\documentclass{ar-1col-S2O}
\usepackage[numbers]{natbib}
\usepackage{url}
\usepackage{booktabs}
\usepackage{amsmath}
\usepackage{amssymb}
\usepackage{caption}
\usepackage{subcaption}
\usepackage{nameref}

\usepackage{hyperref}
\hypersetup{
    colorlinks=true,
    linkcolor=blue,
    citecolor=blue,
    urlcolor=blue
}

\jname{Ann. Rev. Nucl. Part. Sci.}
\jvol{74}
\jyear{2024}
\doi{10.1146/annurev-nucl-102122-022747}

\def\beq{\begin{equation}}
\def\eeq{\end{equation}}
\def\bea{\begin{eqnarray}}
\def\eea{\end{eqnarray}}

\def\lr{\left( }
\def\rr{\right) }
\def\le{\left[ }
\def\re{\right] }

\begin{document}

\markboth{Klasen and Paukkunen}{Nuclear PDFs After the First Decade of LHC Data}

\title{Nuclear PDFs After the First Decade of LHC Data}

\author{Michael Klasen$^1$ and Hannu Paukkunen$^{2,3}$
\affil{$^1$Institute for Theoretical Physics, University of Münster, Münster, Germany; email: michael.klasen@uni-muenster.de}
\affil{$^2$Department of Physics, University of Jyväskylä, Jyväskylä, Finland; email: hannu.paukkunen@jyu.fi}
\affil{$^3$Helsinki Institute of Physics, Helsinki, Finland
}}

\begin{abstract}
We present a review of the conceptual basis, present knowledge and recent progress in the field of global analysis of nuclear parton distribution functions (PDFs). After introducing the theoretical foundations and methodological approaches for the extraction of nuclear PDFs from experimental data, we discuss how different measurements in fixed-target and collider experiments provide increasingly precise constraints on various aspects of nuclear PDFs, including shadowing, antishadowing, the EMC effect, Fermi motion, flavor separation, deuteron binding, target-mass and other higher-twist effects. Particular emphasis is given to measurements carried out in proton-lead collisions at the Large Hadron Collider, which have revolutionized the global analysis during the past decade. These measurements include electroweak-boson, jet, light-hadron, and heavy-flavor observables. Finally, we outline the expected impact of the future Electron Ion Collider and discuss the role and interplay of nuclear PDFs with other branches of nuclear, particle and astroparticle physics. 
\end{abstract}

\begin{keywords}
quantum chromodynamics, nuclear structure, parton distribution functions, collider physics, future developments
\end{keywords}
\maketitle

\tableofcontents

\clearpage
\section{Introduction}
\label{sec:01}

The nuclear structure at high energies is an important current research topic, relevant not only for our understanding of the fundamental quark and gluon dynamics in protons and neutrons bound in nuclei, but also for elucidating the formation, properties and evolution of a deconfined state of hadronic matter that existed in the early Universe, the so-called quark-gluon plasma (QGP). Nuclear parton distribution functions (PDFs) encode cold binding effects in nuclei \cite{Frankfurt:1988nt,Arneodo:1992wf}, determine the pre-equilibrium phase and initial-state phase transition to the QGP \cite{Gelis:2010nm}, and are correlated with the precise extraction of important QGP properties such as its temperature from photons \cite{Klasen:2013mga} and the final-state phase transition during chemical freeze-out \cite{Andronic:2017pug}. While the evolution of the PDFs with the scale $Q^2$, at which they are probed, can be computed in perturbative Quantum Chromodynamics (QCD) \cite{Gribov:1972ri,Gribov:1972rt,Dokshitzer:1977sg,Altarelli:1977zs}, their dependence on the longitudinal parton momentum fraction inside the hadron is non-perturbative and must be fitted to experimental data. Traditionally, deep-inelastic scattering (DIS) of charged leptons or neutrinos and Drell-Yan (DY) dilepton production with fixed targets have provided the bulk of the data. However, over the last decade collider data from the Large Hadron Collider (LHC) at CERN and also from the Relativistic Heavy Ion Collider (RHIC) at BNL have led to significant improvements in our knowledge of collinear, unpolarized nuclear PDFs, which are reviewed here.

In the naive parton model, the double-differential charged-lepton DIS cross section per nucleon
\beq
 \frac{d^2\sigma^{lA}}{dxdy}=\frac{2\pi\alpha^2}{Q^4}s\le1+(1-y)^2\re F_2^A(x)
 \label{eq:01}
\eeq
is directly related to the nuclear structure function $F_2^A(x)=x\sum_q e_q^2f_q^A(x)$ and thus the quark PDFs $f_q^A(x)$ in the nucleus $A$, while the gluon PDF and $Q^2$-dependence enter only in the QCD-improved parton model (see Sec.\ \ref{sec:02}). Here, $\alpha$ is the electromagnetic fine-structure constant, $e_q$ the fractional charge of quark $q$, $Q^2$ the virtuality of the exchanged photon, $s=Q^2/(xy)$ the hadronic center-of-mass energy, $y$ the lepton inelasticity, and $x$ the Bjorken variable. The nuclear structure functions differ from free nucleon structure functions not only due to an admixture of protons and neutrons, but also in various other ways depending on the region in $x$. At small $x$ a depletion or {\em shadowing} is observed in $F_2^A(x)$ at $x\lesssim0.05$, followed with increasing $x$ by an enhancement or {\em antishadowing} at $0.05\lesssim x\lesssim0.3$, again a depletion -- the famous {\em EMC effect} -- at $0.3\lesssim x\lesssim0.7$, and ultimately an enhancement due to {\em Fermi motion} at $0.7\lesssim x$. The ratios of large-$A$ isoscalar structure functions to the structure function of the deuteron (D) -- a loosely bound isoscalar state of a proton ($p$) and a neutron ($n$) -- can be parameterized and fitted, albeit with considerable uncertainties, to SLAC \cite{Arnold:1983mw,Gomez:1993ri} and NMC \cite{NewMuon:1995cua} data in the form \cite{NuTeV:2005wsg}
\beq
 R(x)= 1.10 - 0.36x - 0.28e^{-21.9x} + 2.77x^{14.4},
 \label{eq:02}
\eeq
i.e.\ without a dependence on the nuclear mass number $A$ or the hard scale $Q^2$. However, a logarithmic decrease in $A$ has been observed in the EMC region in the 1984 SLAC data and a logarithmic increase in $Q^2$ in the shadowing region in the more precise NMC data \cite{NewMuon:1996gam}. To first approximation, this is then also how quark PDFs in a bound nucleon $N$ differ from the free-nucleon PDFs.  

The focus of this review is on model-independent global fits of nuclear PDFs and the progress due to LHC data over the last decade. To obtain a first impression of the nuclear dynamics, it is nonetheless interesting to discuss the main historical experimental measurements and theoretical interpretations in the different $x$ regions.

Shadowing had been known to be present in real and virtual photon scattering on nuclei since the 1970s \cite{Caldwell:1978ik,Goodman:1981hc}. At the hadron level, it can be interpreted by assuming that the photon fluctuates from its pointlike state into a superposition of vector mesons ($\rho$, $\omega$, $\phi$), which then interact strongly with the nucleons on the surface of the target nucleus (vector meson dominance). Such nucleons absorb most of the incoming ``hadron'' flux and therefore cast a shadow onto the inner ones \cite{Bauer:1977iq}. At the parton level and in the nuclear rest frame, the photon can be seen to split into a quark-antiquark dipole with lifetime $\tau$, that scatters coherently from multiple partons in the nucleus, if $\tau \geq R_A\sim A^{1/3}$ fm or $x\leq1/(2M_NR_A)\sim0.1A^{-1/3}$, resulting again in a reduced nuclear cross section \cite{Armesto:2006ph,Frankfurt:2011cs,Kopeliovich:2012kw}.

The relative motion of nucleons inside the nucleus was considered first in the 1970s for the deuteron \cite{Atwood:1972zp} and in the 1980s also for heavier nuclei near the Fermi surface \cite{Bodek:1980ar,Frankfurt:1980se,Saito:1985ct}. The structure function of a nucleon in a nucleus can then be expressed as a convolution
\bea
 F_2^A(x_N)=\int_{x_N}^A dy f_A(y) F_2^N\lr \frac{x_N}{y}\rr, &\,{\rm where}\,&
 f_A(y) \sim \frac{1}{\sqrt{2\pi \Delta_A}} \exp{\left\{ \frac{-[y-(1-\delta_A)]^2}{2 \Delta_A^2} \right\} },
 \label{eq:03}
\eea
of the nucleon structure function $F_2^N(x_N)$ with the nucleon momentum distribution $f_A(y)$. Its peak is shifted away from unity due to soft nuclear interactions by an amount $\delta_A\sim0.04$, which corresponds roughly to the ratio of nucleon separation energy over its mass. The width $\Delta_A$ is determined by a fraction of the Fermi momentum $k_F\sim250$ MeV divided by the nucleon mass $M_N\sim1$ GeV and thus small. Eq.\ \ref{eq:03} can therefore be approximated by a simple rescaling $F_2^A(x_N)=F_2^N(x_N/(1-\delta_A))$. The net result of this rescaling is to deplete the partons in the intermediate $x_N$ region, implying $F_2^A/F_2^B<1$ for $A>B$, and to enrich the large $x_N\sim1$ region with $F_2^A/F_2^B>1$. The region $x_N>1$ can be modeled by modifying the Gaussian ansatz for $f_A(y)$ in Eq.\ \ref{eq:03} (e.g.\ with a power law tail \cite{CiofidegliAtti:1995qe}) and also in deconfinement or cluster models \cite{Arneodo:1992wf}, but this region is usually neglected \cite{Segarra:2020gtj}.

The discovery of a suppression of $F_2^{\rm Fe}/F_2^{\rm D}$ at $x_N=0.65$ of $\sim0.89$ in 1983 by the EMC collaboration in muon DIS \cite{EuropeanMuon:1983wih} and its confirmation in reanalyzed iron and aluminum SLAC data from the early 1970s \cite{Bodek:1983qn,Bodek:1983ec} came as a big surprise, since Fermi motion models predicted an enhancement of $\sim1.25$ at this value of $x_N$ and a suppression only for $x_N<0.5$. It triggered many theoretical explanations at both the nuclear and the partonic level, and a consensus still has to emerge \cite{Geesaman:1995yd}. Models with nucleons as the only degrees of freedom in the nucleus must be incomplete, since the convolution in Eq.\ \ref{eq:03} violates baryon number and momentum sum rules. The missing momentum could be carried by pions, whose exchanges lead to an intermediate-range nuclear attraction of $300-500$ MeV, that is canceled by short-distance vector exchanges of almost equal size. The net effect is an average binding energy of 8 MeV per nucleon as observed \cite{Thomas:2018kcx}. However, one would then expect an enhancement of antiquarks and therefore of the DY process, which has not been seen \cite{Alde:1990im}. The failure of nucleon-only and nucleon-pion models indicates that the nucleon structure itself is modified by the medium. The parton model interpretation of the EMC effect is that the medium reduces the number of high-momentum quarks. This momentum reduction leads, via the uncertainty principle, to the notion that quarks in nuclei are confined in a larger volume than that of a free nucleon. There are two proposals to realize this simple idea: either scalar and vector mean-field effects cause bound nucleons to be larger than free ones, or short-range correlations (SRCs) cause the nucleon structure to be modified by including either $NN^*$ configurations or deconfined six-quark configurations that are orthogonal to the two-nucleon wave functions. Interestingly, two-nucleon SRCs might explain the observed linear correlation between the magnitude of the EMC effect at $0.3\leq x_N\leq 0.7$ and the size of the plateau observed in quasi-elastic scattering at $1.5 \leq x_N \leq 2$, which would solve the single-nucleon sum rule problem \cite{Malace:2014uea,Hen:2016kwk}.

A similar compensation mechanism could be at work in the antishadowing region, which is imposed by shadowing through the momentum sum rule. In the Breit frame, small momentum quarks and gluons, because of the uncertainty principle, spread over a distance comparable to the nucleon-nucleon separation. Quarks and gluons from different nucleons can then overlap spatially and fuse, thus increasing the density of high momentum partons (antishadowing) at the expense of that of lower momentum ones (shadowing) \cite{Nikolaev:1975vy}. In perturbative QCD, this process is flavor-dependent and $q\bar{q}\to g$ fusion results, e.g., in shadowing for antiquarks and antishadowing for gluons. The fact that there is no clear evidence of antishadowing in the DY process can be interpreted either with an important role of valence ($v$) quarks or as a consequence of the evolution in $Q^2$ \cite{Frankfurt:1988nt,Arneodo:1992wf}.

The non-perturbative nature of nuclear interactions, need for phenomenological models and incongruous nuclear and partonic interpretations of the effects described above are strong motivations to parameterize and fit nuclear PDFs to the available data in a model-independent way. Improving on Eq.\ \ref{eq:02}, Eskola parameterized in 1992 the ratio of heavier nuclear structure functions over deuterons separately in each region at the starting scale $Q_0^2$, matched it at the transition points (whose definition depended on $A$), and evolved it in $Q^2$ \cite{Eskola:1992zb}. As observed experimentally \cite{Goodman:1981hc} and predicted theoretically \cite{Qiu:1986wh}, shadowing then vanished only very slowly towards larger $Q^2$, in particular for quarks and antiquarks. When the parameterized ratio was fitted to DIS and $pA$ DY data, while imposing baryon number and momentum conservation, the nuclear data could be described fairly independently of the underlying proton PDFs \cite{Eskola:1998df}. A rigorous statistical analysis of DIS data initially led to rather large values of $\chi^2$ per degree of freedom (dof) of $1.82-1.93$ \cite{Hirai:2001np}, which could, however, be reduced to 1.35 in leading order (LO) and 1.21 in next-to-leading order (NLO) QCD using more precise DIS and $pA$ DY data \cite{deFlorian:2003qf,Hirai:2007sx}. Collider (RHIC) data on $\pi^0$ production from 2006 introduced sensitivity to the gluon density beyond scaling violations with a resulting $\chi^2/$dof $=0.79$ in LO and NLO \cite{Eskola:2009uj}. An equally good value of $\chi^2/$dof $=0.83$ was obtained in NLO with a similar data set, but with a direct $A$-dependent parameterization of nuclear PDFs at the starting scale $Q_0^2$ \cite{Kovarik:2015cma}. The inclusion of neutrino DIS proved to be more difficult \cite{Schienbein:2007fs}, which triggered a discussion about the universality of nuclear effects in charged-lepton and neutrino scattering \cite{Paukkunen:2010hb,Kovarik:2010uv,Paukkunen:2013grz,Nakamura:2016cnn,Kalantarians:2017mkj}. In the last decade, a wealth of LHC data on electroweak boson, photon, light and heavy hadron, and jet production have become available for proton-lead ($p$Pb) collisions, which have already had a significant impact on the determination of nuclear PDFs \cite{Duwentaster:2022kpv,Eskola:2021nhw,AbdulKhalek:2022fyi}. These modern developments will be reviewed thoroughly in the following, whereas other recent reviews have mostly focused on proton PDFs \cite{Gao:2017yyd,Kovarik:2019xvh,Ethier:2020way} or sketched a larger multi-dimensional picture of the nucleus \cite{Klein:2019qfb}.

The remainder of this article is organized as follows: in Sec.\ \ref{sec:02}, we briefly review the theoretical foundations of nuclear DIS and its factorization. In Sec.\ \ref{sec:03}, we describe the different methodological approaches in the global fits of nuclear PDFs. In Secs.\ \ref{sec:04} and \ref{sec:05}, we discuss the impact of the different experimental data in roughly chronological order, the main focus being of course on the LHC. The impact of the future Electron Ion Collider and connections to other fields in nuclear, particle and astroparticle physics (lattice QCD, the search for gluon saturation, the QGP and astrophysical phenomena) are briefly addressed in Sec.\ \ref{sec:06}, before we conclude this review in Sec.\ \ref{sec:07}.

\section{Theoretical foundations}
\label{sec:02}

We start our discussion of nuclear PDFs by reviewing their theoretical foundations both within the operator product expansion (OPE) and the QCD-improved parton model, including target-mass and other higher-twist effects.

\subsection{Factorization in the OPE and QCD-improved parton model}
\label{sec:2.1}

\begin{figure}[h]
\centering
\includegraphics[width=\textwidth]{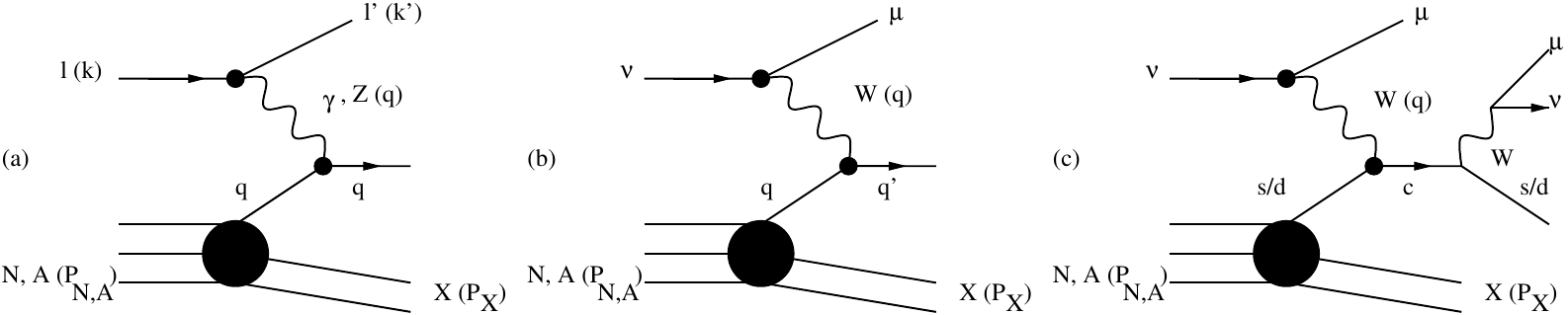}
\caption{Leading-order diagrams for (a) neutral-current (NC) DIS with charged leptons, (b) charged-current (CC) DIS with neutrinos or antineutrinos, and (c) charm dimuon production. The charm quark is understood to hadronize before the semileptonic decay.
}
\label{fig:01}
\end{figure}

Deep-inelastic scattering of high-energy leptons is the key process for studying the hadronic structure of nucleons $N$ or nuclei $A$ with mass $M_N$ or $M_A=A\,M_N$ and four-momentum $P_N$ or $P_A$ in terms of their partonic (quark and gluon) degrees of freedom (cf.\ Fig.\ \ref{fig:01}) \cite{Blumlein:2012bf,Blumlein:2023aso}. The charged lepton $l$ or (anti-)neutrino $\nu$ has incoming (outgoing) four-momentum $k$ ($k'$), the squared center-of-mass energy is $s_{N,A}=(k+P_{N,A})^2$, and $X$ represents all final-state hadrons with total four-momentum $P_X$ and squared mass $W_{N,A}^2=(q+P_{N,A})^2$. In the laboratory frame, the lepton energy loss is $\nu=q \cdot P_{N,A}/M_{N,A}=E-E'$, and the exchanged vector boson $V$ has squared momentum transfer $Q^{2}=-q^{2}>0$.

The inclusive differential cross section $d\tilde{\sigma}^A \sim L^{\mu\nu} \tilde{W}^A_{\mu\nu}$ can be written as a combination of a pointlike leptonic tensor $L^{\mu\nu}$ and the hadronic tensor
\beq
\tilde{W}^A_{\mu\nu}(P_A,q) = \frac{1}{4\pi} \int d^4z ~ e^{iq\cdot z} \langle A \vert J_\mu^\dagger(z)\  J_\nu(0) \vert A\rangle = \frac{1}{4\pi}\ {\rm disc}\, \tilde{T}_{\mu\nu}(P_A,q),
\eeq
where the latter is given in terms of a product of hadronic currents and can be related to the discontinuity of the virtual forward Compton scattering amplitude
\beq
\tilde{T}^A_{\mu\nu}(P_A,q) = \int d^{4}z\ e^{iq\cdot z}\ \langle A\vert \mathcal{T}J_\mu^\dagger(z)\ J_\nu(0)\vert A\rangle.
\eeq
The Operator Product Expansion (OPE) then allows one to expand the hadronic matrix element of the forward scattering amplitude in a complete set of local operators \cite{Georgi:1976ve}
\beq
\tilde{T}^A_{\mu\nu} (P_A,q) = -2i\sum_{j,\tau,n} c_{\tau,\mu\nu}^{j,\mu_1 \cdots \mu_n}
        \langle A | O^{j,\tau}_{\mu_{1} \cdots \mu_{n}} | A \rangle \ =
-2i \sum_{j,k}
\frac{2^{2k} }{ Q^{4k}} C_j^{2k} A_{2k} \tilde{\Pi}^{j,k}_{\mu\nu} \ + \ \mathcal{O}(\tau>2),
\eeq
where $c_{\tau,\mu\nu}^{j,\mu_1 \cdots \mu_n}$ denote the hard scattering, $\tau$ the twist of the operator $O$, defined as its mass dimension minus its spin, and $j$ different operators with the same twist. Up to power corrections, one can identify the product of the perturbative Wilson coefficients $C_i^{2k}$ and reduced hadronic matrix elements $A_{2k}$ as integer Mellin moments of structure functions
\beq
\int_0^1 dy ~ y^{2k-1} ~ \tilde{F}^A_i(y,Q^2) = C_i^{2k} A_{2k} + \mathcal{O}(\tau>2) \,,
\label{eq:04}
\eeq
with $y^{2k-1} \to y^{2k-2}$ for $i=2$. The Lorentz structure in terms of metric tensors and momenta is encoded in $\tilde{\Pi}^{j,k}_{\mu\nu}$.

In the QCD-improved parton model, the nuclear structure functions
\begin{equation}
    {\tilde{F}}^A_i(x_A,Q^2)
    = \sum_{j=q,g} \int_{x_A}^1 \frac{dy_A}{y_A} \ C_{i,j} \tilde{f}_j^A(y_A,Q^2) \ + \mathcal{O}(\tau>2)
\label{eq:05}
\end{equation}
depend on the Bjorken scaling variable $x_{N,A}=Q^{2}/(2q \cdot P_{N,A})=Q^{2}/(2\nu M_{N,A})$ with $x_N\in[0,A]$ ($x_A=x_N/A\in[0, 1]$) with logarithmic scaling violation in $Q^2$ (see below). They are given as convolutions of target-independent short-distance Wilson coefficients $C_{i,j}$ with universal nuclear parton distribution functions (PDFs) $\tilde{f}_j^A$. Inspection of Eqs.\ \ref{eq:04} and \ref{eq:05} shows that nuclear PDFs can be understood as moments of matrix elements of local twist-two operators composed of quark and gluon fields \cite{Collins:1989gx}. The nuclear PDFs $\tilde{f}_i^A(x_A,Q^2)$ above are related to the more familiar average-nucleon nuclear PDFs $f_i^A(x_N,Q^2)$ through $f_i^A(x_N,Q^2) = \tilde f_i^A(x_A,Q^2)/A$. This rescaling is a key step that allows us to compare structure functions across different nuclei, including the free nucleon. The evolution of the PDFs with the scale $Q^2$ is perturbatively calculable and given by the Dokshitzer-Gribov-Lipatov-Altarelli-Parisi (DGLAP) evolution equations 
\cite{Gribov:1972ri,Gribov:1972rt,Dokshitzer:1977sg,Altarelli:1977zs}
\begin{eqnarray}
\frac{df_i^A(x_N,Q^2)}{d \ln Q^2} & = &
\frac{\alpha_s(Q^2)}{2 \pi} \int_{x_N}^A \frac{dy_N}{y_N}\
P_{ij}\left(\frac{x_N}{y_N} \right)  \  f_j^A(y_N,Q^2) \,,
\end{eqnarray}
where $\alpha_s$ is the QCD coupling and $P_{ij}$ are the partonic splitting functions. Furthermore, the PDFs satisfy sum rules due to charge, baryon number and momentum conservation,
\bea
 \int_0^A dx_N \{f_{u_v}^A,f_{d_v}^A\}(x_N,Q^2) = \{2Z+N,Z+2N\} &,&
 \int_0^A dx_N x_N \sum_i f_i^A(x_N,Q^2) = A,
\eea
where $Z$ is the electric charge of the nucleus with baryon number $A=Z+N$. It is therefore common (but not necessary) to decompose the nuclear PDFs as
\begin{equation}
    f_i^{A}(x_N,Q^2) =
    \frac{Z}{A} f_i^{p/A}(x_N,Q^2)
    + \frac{A-Z}{A} f_i^{n/A}(x_N,Q^2)
    \,,
 \label{eq:11}
\end{equation}
where the bound neutron PDFs $f_i^{n/A}(x_N,Q^2)$ are commonly obtained from those of the bound proton $f_i^{p/A}(x_N,Q^2)$ by assuming isospin symmetry,
\begin{equation}
f_{u, \bar{u}}^{n/A}(x_N,Q^2) = f_{d, \bar{d}}^{p/A}(x_N,Q^2) \,, \ \
f_{d, \bar{d}}^{n/A}(x_N,Q^2) = f_{u, \bar{u}}^{p/A}(x_N,Q^2) \,.
\end{equation}
In principle the above integrations extend to $A$, although the dominant support of the PDFs is expected to be in the region $x_A\leq 1/A$, or $x_N\leq 1$. One therefore usually assumes $f_i^A(x_N,Q^2)=0$ for $x_N>1$, which has the advantage that the same evolution equations can be used for all nuclei in the interval $x\in[0,1]$. 

Accounting for the heavy-quark masses is essential in an accurate description of the free proton data \cite{Nadolsky:2009ge}. Standard methods to handle the quark masses are nowadays General Mass (GM) Variable Flavor Number Schemes (VFNS) \cite{Thorne:2008xf} such as the simplified Aivazis-Collins-Olness-Tung (SACOT) schemes \cite{Aivazis:1993pi,Kramer:2000hn,Tung:2001mv} or the FONLL schemes \cite{Cacciari:1998it,Forte:2010ta}, which provide systematic ways to interpolate between the Fixed Flavor Number Scheme (FFNS), in which heavy quarks are not considered as partons, and the Zero Mass (ZM) VFNS, in which heavy quarks are treated as massless partons.

\subsection{Target mass corrections}
\label{sec:2.2}

Target mass corrections (TMCs) can be discussed in terms of collinear factorization \cite{Accardi:2008ne} or be obtained from the OPE by inverting moments of structure functions, cf.\ Eq.\ \ref{eq:04} \cite{Georgi:1976ve}. They can be written in the general form \cite{Ruiz:2023ozv}
\beq
F_i^{A,{\rm TMC}}(x_N,Q^2)=\sum_j A_i^j F_j^A(\xi_N,Q^2) + B_i^j h_j^A(\xi_N,Q^2) + C_i g_2^A(\xi_N,Q^2),
\eeq
where $\xi_N\!=\!2x_N/(1+r_N)$ is the Nachtmann variable \cite{Nachtmann:1973mr} with $r_N\!=\!\sqrt{1+4x_N^2M_N^2/Q^2}$ and
\bea
h_j^A(\xi_N,Q^2)\sim\int_{\xi_N}^A d\xi_N' \frac{F_j^{A}(\xi_N',Q^2)}{\xi_N'}&\quad ,\quad &
g_2^A(\xi_N,Q^2)=\int_{\xi_N}^A d\xi_N' h_2^A(\xi_N',Q^2)
\eea
are auxiliary functions with $1/\xi_N'\to1/{\xi_N'}^2$ for $j=2$. Specifically, the proportionality factor in $h_2^A$ is unity and we have
\beq
F_{2}^{A,{\rm TMC}}(x_N,Q^{2})  =  \left(\frac{x_N^{2}}{\xi_N^{2}r_N^{3}}\right) F_{2}^{A}(\xi_N,Q^{2}) +\left(\frac{6 M_N^2 x_N^{3}}{Q^2 r_N^{4}}\right) h_{2}^A(\xi_N,Q^{2}) +\left(\frac{12 M_N^4 x_N^{4}}{Q^4 r_N^{5}}\right) g_{2}^A(\xi_N,Q^{2}).
\eeq
Quark masses modify $\xi_N$ by $\xi_N \rightarrow R_{ij} \xi_N$, where the factor $R_{ij}$ depends on the incoming and outgoing quark masses $m_{i,j}$. In the case $m_i=0$, one obtains the slow-rescaling limit $R_{ij}=1+(n m_j)^2/Q^2$ with $n=1$ for CC \cite{Barnett:1976ak} and $n=2$ for NC DIS \cite{Nadolsky:2009ge}.

\subsection{Higher twist corrections}
\label{sec:2.3}

\begin{figure}[h]
\centering
\begin{subfigure}[b]{0.31 \textwidth}
\includegraphics[width=\textwidth]{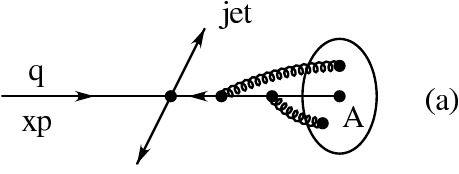}
\end{subfigure}
\hfill
\begin{subfigure}[b]{0.31\textwidth}
\includegraphics[width=\textwidth]{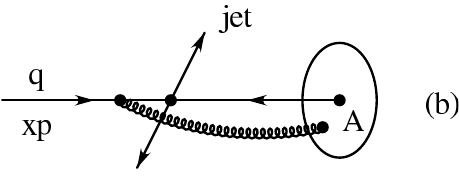}
\end{subfigure}
\hfill
\begin{subfigure}[b]{0.31\textwidth}
\includegraphics[width=\textwidth]{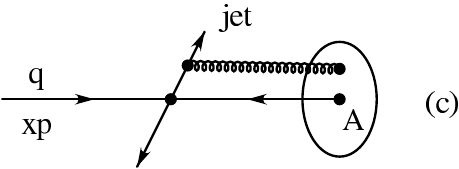}
\end{subfigure}
\caption{Classification of multiple parton scattering in nuclear medium: (a) interactions internal to the nucleus, (b) initial-state (IS) and (c) final-state (FS) interactions \cite{Accardi:2004be}.}
\label{fig:02}
\end{figure}

Interactions internal to the nucleus as in Fig.~\ref{fig:02} (a) change the nuclear PDFs with respect to those of the free nucleon. However, since only a single parton participates in the hard scattering, the structure functions can still be factorized as in Eq.\ \ref{eq:05}, and the leading twist nuclear PDFs can be parameterized at an initial scale $Q_0$, evolved with (in principle $A$-dependent) evolution equations and fitted to experimental data or modeled theoretically.

Next-to-leading power corrections to Eq.\ \ref{eq:05} of ${\cal O}(r_T^2\sim1/p_T^2)$, ${\cal O}(m_J^2/p_T^2)$ and ${\cal O}(\alpha_s(Q^2)\Lambda^2/Q^2)$ arise from the transverse size $r_T$ of the initial nucleus, the non-vanishing invariant mass $m_J$ of the final jet, and IS and FS interactions involving more than one parton as shown in Figs.~\ref{fig:02} (b) and (c). In hadron-nucleus collisions, both are enhanced by $A^{1/3}$ due to the large density of soft partons in the nucleus. For IS interactions, $\Lambda^2\sim 0.01$ GeV$^2$ is the squared scale of the twist-four correlation function of single parton pairs and proportional to the transverse field strength. For FS interactions, long-range soft parton interactions must also be considered. It can be shown that the $A^{1/3}$ enhancement can be factorized to all powers in hadron-nucleus, but not nucleus-nucleus collisions and then involves correlation functions of multiple parton pairs. In general, however, even the hadron-nucleus Drell-Yan cross section cannot be factorized beyond next-to-leading power \cite{Accardi:2004be}.

\section{Global analyses of nuclear PDFs}
\label{sec:03}

The inverse problem of extracting nuclear PDFs from experimental data is approached in a similar way as the determination of (free) proton PDFs \cite{Ethier:2020way}, i.e.\ all global analyses are based on optimizing the correspondence between theoretical calculations and experimental measurements by minimizing a figure-of-merit function that is typically of the form
\begin{equation}
\chi^2 = \sum_{i,j} \left(D_i - T_i \right) C^{-1}_{ij} \left(D_j - T_j \right) \,.
\end{equation}
Here, $D_i$ denote the experimental values for observables in the fit, and $T_i$ are the corresponding theoretical values, which depend on the PDFs. The covariance matrix is defined as $C_{ij} = \sigma_i^2\delta_{ij}+\sum_\alpha\bar{\sigma}_{i\alpha}\bar{\sigma}_{j\alpha}$, where $\sigma_{i}$ is the total uncorrelated uncertainty added in quadrature and $\bar{\sigma}_{i\alpha}$ is the correlated systematic uncertainty from source $\alpha$. How to exactly assign values for $\sigma_{i}$ and $\bar{\sigma}_{i\alpha}$ varies from one analysis to another and depends on whether uncertainties are multiplicative or additive \cite{DAgostini:1993arp,Ball:2009qv}. In the case of fitting nuclear PDFs, only a handful of the data sets provide the full information on the correlated systematic uncertainties -- in most of the cases only the overall normalization uncertainty is given.

An essential part of global PDF analyses is the propagation of experimental uncertainties into the PDFs. From the practical point of view, the two principal methods are the Hessian \cite{Pumplin:2000vx,Pumplin:2001ct} and Monte Carlo methods \cite{Giele:1998gw,Giele:2001mr}. The Hessian uncertainty analysis \cite{Pumplin:2000vx,Pumplin:2001ct} is based on expanding the $\chi^2$ function in the vicinity of its minimum value $\chi^2_0$,
\begin{equation}
\chi^2 \approx \chi^2_0 + \sum_{i,j} H_{ij} \delta a_i \delta a_j = \chi^2_0 + \sum_{i} z_i^2 \,,
\end{equation}
where $\delta a_i$ are deviations from the best-fit parameters and $H_{ij}$ is the second-derivative matrix, the Hessian matrix. In the second step one diagonalizes the Hessian matrix by finding its eigendirections. The PDF error sets $S_k^\pm$ are then defined as deviations along these eigendirections to positive/negative directions such that $\chi^2$ increases by a fixed amount $\Delta\chi^2$. The value for $\Delta\chi^2$ can be defined in various ways. A common feature in the current global analyses is that $\Delta\chi^2$ is of the order of the number of fit parameters for 68\% confidence-level (CL) uncertainties and somewhat higher for 90\% CL uncertainties. In the plots of this review, the uncertainties $\big( \delta X \big)^\pm$ for a given PDF-dependent quantity $X$ are calculated by the asymmetric prescription \cite{Nadolsky:2001yg}, 
\begin{align}
\big( \delta X \big)^\pm \label{eq:errorprop2} 
& = 
\sqrt{
\sum_k \bigg[ \substack{ \max \\ \min}  \left[X(S_k^+)-X(S_0), X(S_k^-)-X(S_0), 0\right]\bigg]^2
} 
\,,
\end{align}
where $S_0$ denotes the best fit. Out of the fits discussed in this review, nCTEQ15HQ \cite{Duwentaster:2022kpv}, EPPS21 \cite{Eskola:2021nhw}, TUJU21 \cite{Helenius:2021tof} and KSASG20 \cite{Khanpour:2020zyu} make use of the Hessian method. The Monte Carlo method is based on preparing several fits in which the central values of the experimental data have been randomly shifted within the uncertainties. In the case of uncorrelated experimental uncertainties, the fitted data points are obtained through
\begin{equation}
D_i \rightarrow D_i \left( 1 + \sigma_i R_i \right) \,, 
\end{equation}
where $R_i$ is a taken from a Gaussian distribution centered around $0$ and with a unit standard deviation. The above formula can also be generalized to the case of correlated uncertainties \cite{Forte:2002fg,Ball:2008by}. In the nNNPDF3.0 fit \cite{AbdulKhalek:2022fyi}, the 100\% uncertainty band for a given quantity $X$ is then defined as the minimum/maximum value obtained by calculating $X$ with all PDF replicas. The 90\% uncertainty is defined by disregarding the highest/lowest 10\%. 

\subsection{The nCTEQ framework}
\label{sec:3.1}

In the nCTEQ15 NLO analysis \cite{Kovarik:2015cma} and its sequels, the ansatz for the nuclear PDFs at $Q_0 = 1.3$ GeV follows the CTEQ6M parameterization \cite{Pumplin:2002vw}
\beq
 xf_i^{p/A}(x,Q_0^2)=c_{0i}x^{c_{1i}}(1-x)^{c_{2i}}e^{c_{3i}x}(1+e^{c_{4i}}x)^{c_{5i}},
\eeq
where $i=u_{v},d_{v},g,\bar{u}+\bar{d},s+\bar{s}$ and we have dropped the index $N$ of $x$, while
\beq
\frac{f^{p/A}_{\bar{d}}(x,Q_{0}^2)}{f^{p/A}_{\bar{u}}(x,Q_{0}^2)}
= c'_{0}\,x^{c'_{1}}(1-x)^{c'_{2}}+(1+c'_{3}x)(1-x)^{c'_{4}}.
\eeq
The normalization coefficients $c_{0i}$ are constrained by the momentum and valence quark sum rules. The proton baseline is similar to the fit CTEQ6.1M \cite{Stump:2003yu}, but has minimal influence from nuclear data \cite{Owens:2007kp}. There are currently no uncertainties associated with this proton baseline PDF. The $A$-dependence of the parameterization is directly included in the coefficients
\beq
 c_{ki} \longrightarrow c_{ki}(A) \equiv p_{ki}+ a_{ki}(1-A^{-b_{ki}}), \quad k = \{1, ..., 5\}.
\eeq
The 16 free parameters in nCTEQ15 describe the $x$-dependence in $u_v$, $d_v$, $g$, $\bar{u} + \bar{d}$, while the parameters in $\bar{d}/\bar{u}$ were fixed, as was $s=\bar{s}=\kappa(\bar{u}+\bar{d})/2$. Starting with the fits nCTEQ15WZ \cite{Kusina:2020lyz} and nCTEQ15WZ+SIH \cite{Duwentaster:2021ioo} including weak ($W,Z$) boson and single-inclusive hadron (SIH) production at the LHC, three free parameters were added for $s+\bar{s}$. The 38 error PDFs are obtained with the Hessian method and a tolerance of $\Delta\chi^2=35$.

In the latest fit nCTEQ15HQ \cite{Duwentaster:2022kpv}, the weakly fragmentation-function dependent SIH data were complemented by open heavy quark and quarkonium (HQ) production data from the LHC that had shown great potential to constrain the gluon in preceding reweighting studies \cite{Eskola:2019bgf,Kusina:2020dki}. These data were fitted with a data-driven method \cite{Kom:2011bd}, in which the cross sections of hadrons 
$A_1$ and $A_2$ are taken to be dominated by the gluon-gluon subprocesses,
\beq
\sigma(A_1 A_2 \rightarrow \mathcal{Q}+X) = \int d x_{1} d x_{2} f_g^{A_1}\left(x_{1}, \mu^2\right) f_g^{A_2}\left(x_{2}, \mu^2\right) \frac{1}{2 \hat{s}} \overline{\left|\mathcal{A}_{g g \rightarrow \mathcal{Q}+X}\right|^{2}} \mathrm{dPS} \,, \label{eq:MEF}
\eeq
where $\mathcal{Q}=D^0,J/\psi,B\!\to\! J/\psi,\Upsilon,\psi',B\!\to\!\psi'$, the squared factorization scale $\mu^2$ is related to the geometric mean of $M_\mathcal{Q}^2$ and $p_T^2$, and $\mathrm{dPS}$ denotes the two-particle phase space. The effective matrix elements $\overline{\left|\mathcal{A}_{g g \rightarrow \mathcal{Q}+X}\right|^{2}}$ are parameterized by a generalized Crystal Ball function,
\begin{align}
\begin{split}
\overline{\left|\mathcal{A}_{g g \rightarrow \mathcal{Q}+X}\right|^{2}}= \frac{\lambda^2\kappa\hat{s}}{M_\mathcal{Q}^2} e^{a|y|}
&\times\begin{cases} e ^{ -\kappa \frac{p_{\rm T}^{2}}{M_{\mathcal{Q}}^{2}}} & \text { if } p_{\rm T} \leq\left\langle p_{\rm T}\right\rangle \\ e^{-\kappa \frac{\left\langle p_{\rm T}\right\rangle^{2}}{M_{\mathcal{Q}}^{2}}}\left(1+\frac{\kappa}{n} \frac{p_{\rm T}^{2}-\left\langle p_{\rm T}\right\rangle^{2}}{M_{\mathcal{Q}}^{2}}\right)^{-n} & \text { if } p_{\rm T}>\left\langle p_{\rm T}\right\rangle\end{cases} , \label{eqn:CrystalBall}
\end{split}
\end{align}
where $M_{\mathcal{Q}}$ denotes the mass of particle $\mathcal{Q}$, $\hat s = x_1 x_2 s$, and $p_{\rm T}$ and $y$ correspond to the transverse momentum and rapidity of $\mathcal{Q}$. The free parameters $\lambda$, $\kappa$, $\left<p_T \right>$, $n$ and $a$ are fitted for each final state to $pp$ data. The fits agree with NLO GM-VFNS \cite{Kniehl:2004fy} and NRQCD \cite{Butenschoen:2010rq} calculations within their scale uncertainties. In the fit nCTEQ15HQ, the LHC $p$Pb data are consistently included in the fit as absolute cross sections in the case of all observables. 

\subsection{The EPPS framework}
\label{sec:3.2}

The latest EPPS analysis, EPPS21~\cite{Eskola:2021nhw}, is rooted in a series of global fits \cite{Eskola:2009uj,Eskola:1998iy,Eskola:2007my,Eskola:2008ca,Eskola:2016oht}, which parameterize the bound proton PDFs at the starting scale $Q_0 = 1.3$ GeV as
\beq
f^{p/A}_i(x,Q_0^2) = R^{p/A}_i(x,Q_0^2) f^p_i(x,Q_0^2) \,. \label{eq:defnPDFEPPS}
\eeq
Here, $i=u_v,d_v,g,\bar{u},\bar{d},s=\bar{s}$, and the free proton PDFs $f^p_i(x,Q^2)$  are taken from the fit CT18A \cite{Hou:2019efy}, which includes more LHC $pp$ data sensitive to strange quarks than the default CT18 fit. This reduces the dependence of the proton baseline on the $\nu$Fe DIS data, which is also part of the CT18A analysis. The nuclear modifications $R^{p/A}_i(x,Q_0^2)$ are parameterized through 24 free parameters. The parameterization is piecewise smooth in $x$, so that parameters controlling different $x$ regions mix as little as possible, i.e.\
\begin{align}
& R_i^{p/A}(x,Q^2_0) = \label{eq:FitFormEPPS21} 
 \left\{
\begin{array}{lr}
a_{0i} + a_{1i}\big(x-x_{ai}\big) 
\Big[e^{-xa_{2i}/x_{ai}}-e^{-a_{2i}} \Big]
\,, &   x \leq x_{ai} \\[5pt]
b_{0i}x^{b_{1i}}\big(1-x\big)^{b_{2i}}
e^{xb_{3i}} \,, 
& \hspace{-0.0cm} x_{ai} \leq x \leq x_{ei} \\[5pt]
c_{0i} + c_{1i}\left(c_{2i}-x \right) \left(1-x\right)^{-\beta_i} \,, & \hspace{-0.0cm} x_{ei} \leq x \leq 1 \,,
\end{array}
\right.  
\end{align}
where $x_{ai}$ and $x_{ei}$ are the locations of the anticipated antishadowing maximum and minimum of the EMC effect, respectively. The $A$-dependence is encoded in such a way that larger nuclei tend to have larger nuclear effects at $x=0, x_{ai}, x_{ei}$ through
\begin{align}
R_i^{p/A}(x,Q^2_0) = 1 + \Big[ R_i^{p/A_{\rm ref}}(x,Q^2_0) - 1 \Big] \left(\frac{A}{A_{\rm ref}} \right)^{\gamma_i} \,, \ \gamma_i > 0 \,, \ A_{\rm ref} = 12 \,. 
\end{align}
However, for very small nuclei a monotonic $A$ scaling is not necessarily a justified assumption (e.g.\ certain small nuclei are more tightly bound), and such deviations are also allowed in the EPPS parameterization through $R^{p/A}_i(x,Q_0^2) \longrightarrow 1 + f_A \Big[R^{p/A}_i(x,Q_0^2) - 1 \Big]$, where $f_A=1$ by default. The parameterization is applied for $A \geq 3$, while for smaller nuclei the nuclear modification is set to unity. This is in line with the baseline CT18A proton PDFs \cite{Hou:2019efy} which include DIS data on deuteron targets with no nuclear corrections \cite{Accardi:2021ysh}. The parameterization of $R^{p/A}_i(x,Q_0^2)$, and thereby the nuclear PDFs, are not restricted to be positive at small $x$. Even if the positivity was imposed at $Q_0$, the backward evolution to smaller $Q^2$ would result in negative values, especially for gluons. Requiring the positivity at $Q_0$ appears thus too restrictive and would induce an increased dependence on the parameterization scale $Q_0$.

The EPPS analyses use ratios of cross sections or structure functions whenever possible in order to remove the dependence on free proton PDFs as much as possible. Ratios of cross sections have the additional advantage that they are perturbatively much more stable than absolute cross sections, reducing the risk of fitting missing higher order effects into the nuclear modifications. Furthermore, experimental uncertainties -- known and unknown -- can be expected to cancel, e.g.\ the one from the luminosity. The uncertainties in the EPPS21 analysis are evaluated through the Hessian method with a global tolerance of $\Delta\chi^2=33$. In addition, the dependence of nuclear modifications on the free proton PDFs is mapped by repeating the fit with each of the 58 CT18A error sets as well. As a result, the EPPS21 fit comes with 106 error sets, which are correlated with the CT18A error sets. 

\subsection{The nNNPDF framework}
\label{sec:3.3}

In the nNNPDF3.0 NLO analysis \cite{AbdulKhalek:2022fyi}, six independent combinations of nuclear PDFs are parameterized at $Q_0=1 \, {\rm GeV}$ in an evolution basis,
\bea
xf_i^{p/A}(x,Q_0^2) & = & B_i x^{\alpha_i} (1-x)^{\beta_i} {\rm NN}_i(x,A) \,, \ i=\Sigma,T_3,T_8,V,V_3,g \,, \label{eq:param2}
\eea
where $\Sigma,T_3,T_8,V,V_3$ label certain combinations of quark PDFs \cite{Gao:2017yyd}, $g$ is the gluon and ${\rm NN}_f(x,A)$ represents the value of the neuron in the output layer of the neural network associated to each independent nuclear PDF. The normalisation coefficients $B_{\Sigma} = B_{T_3} = B_{T_8} = 1$, while $B_V$, $B_{V_3}$ and $B_g$ enforce the momentum and valence sum rules and are determined at $Q_0$ for each value of $A$. The preprocessing exponents $\alpha_i$ and $\beta_i$ are required to control the small and large $x$ behavior of the nuclear PDFs. They are fitted simultaneously with the network parameters. The exponents $\alpha_V$ and $\alpha_{V_3}$ are restricted to lie in the range $[0,5]$ during the fit to ensure integrability of the valence distributions. The other exponents $\alpha_i$ are restricted to the range $[-1,5]$, consistent with momentum sum rule requirements, while the exponents $\beta_i$ lie in the range $[1,10]$.  The figure of merit is defined as
\beq
\label{eq:chi2_fit}
\chi_{\rm fit}^2 = \chi_{\rm t_0}^2 + \kappa_{\rm pos}^2 + \kappa_{\rm BC}^2 \, ,
\eeq
where the first term is the contribution from experimental data with a covariance matrix that takes into account the normalization uncertainties of the different data sets with a self-consistent iterative $t_0$ procedure \cite{Ball:2009qv}. The second term $\kappa_{\rm pos}^2$ imposes the positivity of physical cross sections, and the third term $\kappa_{\rm BC}^2$ ensures that in the limit $A \to 1$ the nNNPDF predictions reduce to those of the free-proton boundary condition using a grid of 100 points, half of which are distributed logarithmically between $x=10^{-6}$ and $0.1$ and the remaining half are linearly distributed between $0.1$ and $0.7$. Ratios of open heavy quark production (i.e.\ $D^0$ meson) data are included with the reweighting method.

The free proton baseline PDFs are fitted separately with the NNPDF3.1 methodology \cite{NNPDF:2017mvq} to all data sets included in the NNPDF4.0 NLO analysis \cite{NNPDF:2021njg} except those involving nuclei with $A \geq$ 2. Each replica of nNNPDF3.0 has been fitted using a randomly chosen replica of proton PDFs coming from these fits. As a result, the uncertainties of nNNPDF3.0 reduce to those of the proton baseline in the $A \to 1$ limit, and the correlations between free-proton and nuclear PDFs are charted and available for the users. The $A \geq$ 2 data that was removed from the proton fit are included in the fit of nuclear PDFs. Consequently, the nuclear effects in deuteron are fitted in a model-independent way.

A common challenge in training neural network models is the choice of the hyperparameters such as their architecture and activation functions, the optimization algorithm and learning rates. Here, Kernel Density Estimators \cite{Rosenblatt:1956,Parzen:1962} have been proven to outperform random or grid searches in selecting the most promising sets. One observes, e.g., that a network with one hidden layer and $25$ nodes beats a network with two hidden layers.

\subsection{NNLO and model-dependent approaches}
\label{sec:3.4}

A few analyses have been performed in next-to-next-to-leading order (NNLO) QCD, albeit with a restricted data set. For example, the nNNPDF1.0 analysis \cite{AbdulKhalek:2019mzd} fitted only the light-quark singlet, octet and gluon PDFs to the charged-lepton DIS data. The TUJU19 analysis \cite{Walt:2019slu} took a step further by complementing the charged-lepton DIS data with neutrino DIS data in the framework of the open source tool xFitter \cite{xFitterDevelopersTeam:2017xal}. Using an ansatz
\begin{equation}
xf^{p/A}_i\left(x,Q_0^2 \right) = c_{0i}\,x^{c_{1i}} (1-x)^{c_{2i}} \left(1+c_{3i}\,x + c_{4i}\,x^2 \right)
\label{pdf-parameterization}
\end{equation}
and a definition of $\chi^2$ that followed the HERAPDF2.0 analysis \cite{H1:2015ubc}, the free proton baseline was first fitted to the HERA data. Due to the limited data set, $s=\bar{s}=\bar{u}=\bar{d}$ had to be assumed, but with 13 free parameters and a tolerance of $\Delta\chi^2=20$, a very similar proton baseline as HERAPDF2.0 was obtained. These proton PDFs were then utilized in the fit to the heavy nuclei data, in which the $A$-dependence was directly encoded in the fit parameters $c_k$ as in the nCTEQ fits \cite{Kovarik:2015cma}, including the case of the deuteron. In the updated TUJU21 analysis \cite{Helenius:2021tof}, also LHC data for weak boson production were included both in the proton and nuclear fits. However, it was still not possible to consider the flavor decomposition in the sea quark sector. The TUJU19 and TUJU21 analyses were the first to consider both the proton and heavier nuclei in the same framework, even though they did not provide a common Hessian matrix. 

The KSASG20 analysis \cite{Khanpour:2020zyu} is based on CT18 \cite{Hou:2019efy} free-proton PDFs and 
uses a nuclear modification factor of cubic HKN form \cite{Hirai:2001np,Hirai:2007sx}
\beq
{\cal W}_i(x, A) =1+\left(1-\frac{1}{A^{\alpha}}\right) \frac{a_{i} (A)+ b_{i}(A) \, x+ c_i(A) \, x^{2}+ d_{i}(A) \, x^{3} }{(1-x)^{\beta_{i}}},
\eeq
which is flexible enough to accomodate both shadowing and antishadowing effects. From nuclear volume and surface contributions, one has $\alpha=1/3$ \cite{Sick:1992pw}, the $a_i$ control shadowing, and $\beta_i$ are related to Fermi motion. Only charged-lepton and neutrino DIS data, supplemented with fixed-target DY data, were fitted with a very restricted flavor decomposition.

Model-dependent nuclear PDFs have been proposed based on the leading twist approximation for nuclear shadowing with antishadowing constrained by the momentum sum rule \cite{Frankfurt:2011cs}, SRCs of nucleon pairs motivated by the EMC effect \cite{Denniston:2023dwd}, and the four-component Kulagin-Petti model \cite{Kulagin:2004ie,Kulagin:2007ju}, which can reproduce a variety of DIS, fixed-target DY \cite{Kulagin:2014vsa} and LHC weak-boson data \cite{Ru:2016wfx}. As the viewpoint of the present review is a data-based global analysis of nuclear PDFs, we will not discuss the modeling of nuclear effects further. 

\subsection{Overview of global nuclear PDF analyses}
\label{sec:3.5}

\begin{figure}[ht]
\centering
\includegraphics[width=0.63\textwidth]{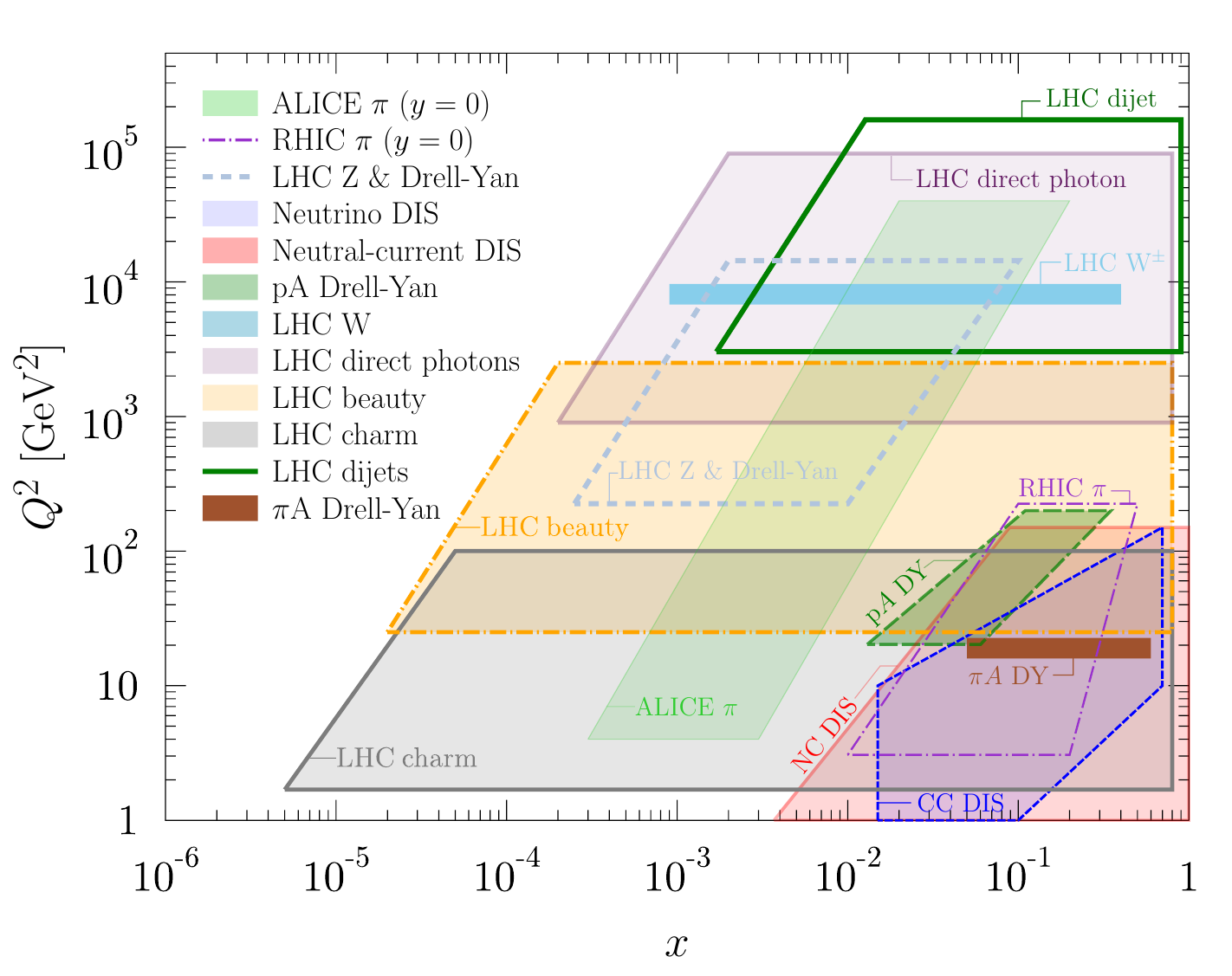}
\caption{An illustration of the $x$ and $Q^2$ regions probed by the current lepton-$A$, pion-$A$ and proton-$A$ data included in the global analyses of nuclear PDFs.}
\label{fig:xQ2plot}
\end{figure}

The key features of the nuclear PDF analyses described above are summarized in Tab.\ \ref{tab:01}. In addition to the main methodological assumptions and input parameters, we also list the fitted experimental data types, grouped into fixed-target and collider data. Since TUJU21 and KSASG20 include (almost) no collider data, we focus on nCTEQ15HQ, EPPS21 and nNNPDF3.0 in the following. Figure \ref{fig:xQ2plot} shows the regions of $x$ and $Q^2$ covered by the world data and included in these analyses. As can be appreciated from the plot, the LHC data taken during the first decade of $pA$ runs have radically expanded the available range both in $x$ and $Q^2$ as well as diversified the global analysis. With this large coverage, the question of process independence of nuclear PDFs can now be addressed much more convincingly.

\renewcommand{\arraystretch}{1.05}
\begin{table}
\tabcolsep4.0pt
\caption{Key features of recent global analyses of nuclear PDFs.}
\label{tab:01}
\hspace{-2.4cm}\begin{tabular}{lcccccc}
\\
\toprule
{\sc Analysis} 
& nCTEQ15HQ~\cite{Duwentaster:2022kpv}
& EPPS21~\cite{Eskola:2021nhw} 
& nNNPDF3.0~\cite{AbdulKhalek:2022fyi}
& TUJU21~\cite{Helenius:2021tof}
& KSASG20~\cite{Khanpour:2020zyu}
\\
\midrule
\textcolor{blue}{\sc Theoretical input:}\ \\ 
  Perturbative order 
& NLO
& NLO
& NLO
& NNLO
& NNLO
\\
  Heavy-quark scheme
& SACOT$-\chi$
& SACOT$-\chi$
& FONLL
& FONLL
& FONLL
\\
  Value of $\alpha_s(M_Z)$
& 0.118
& 0.118
& 0.118
& 0.118
& 0.118
\\
Charm mass $m_c$
& $1.3\,{\rm GeV}$
& $1.3\,{\rm GeV}$
& $1.51\,{\rm GeV}$
& $1.43\,{\rm GeV}$
& $1.3\,{\rm GeV}$
\\
Bottom mass $m_b$
& $4.5\,{\rm GeV}$
& $4.75\,{\rm GeV}$
& $4.92\,{\rm GeV}$
& $4.5\,{\rm GeV}$
& $4.75\,{\rm GeV}$
\\
Input scale $Q_0$
& 1.3\,GeV
& 1.3\,GeV
& 1.0\,GeV
& 1.3\,GeV
& 1.3\,GeV
\\
Data points 
& 1484
& 2077
& 2188
& 2410
& 4353
\\
Independent flavors
& 5
& 6
& 6
& 4
& 3
\\
Parameterization
& Analytic
& Analytic
& Neural network
& Analytic
& Analytic
\\
Free parameters
& 19
& 24
& 256
& 16
& 18
\\
Error analysis
& Hessian
& Hessian
& Monte Carlo
& Hessian
& Hessian
\\
Tolerance
& $\Delta\chi^2=35$
& $\Delta\chi^2=33$
& N/A
& $\Delta\chi^2=50$
& $\Delta\chi^2=20$
\\
Proton PDF
& $\sim$CTEQ6.1
& CT18A
& $\sim$NNPDF4.0
& $\sim$HERAPDF2.0
& CT18
\\
Proton PDF correlations
& 
& \checkmark
& \checkmark
& 
\\
Deuteron corrections
& (\checkmark)$^{a,b}$
& \checkmark$^c$
& \checkmark
& \checkmark
& \checkmark
\\
\midrule
\textcolor{blue}{\sc Fixed-target data:}\ \\
SLAC/EMC/NMC NC DIS
& \checkmark
& \checkmark
& \checkmark
& \checkmark
& \checkmark
\\
-- Cut on $Q^2$ 
& 4 GeV$^2$
& 1.69 GeV$^2$
& 3.5 GeV$^2$
& 3.5 GeV$^2$
& 1.2 GeV$^2$ 
\\
-- Cut on $W^2$
& 12.25 GeV$^2$
& 3.24 GeV$^2$
& 12.5 GeV$^2$
& 12.0 GeV$^2$
\\
JLab NC DIS
& (\checkmark)$^a$
& \checkmark
& 
& 
& \checkmark
\\
CHORUS/CDHSW CC DIS
& (\checkmark/-)$^b$
& \checkmark/-
& \checkmark/-
& \checkmark/\checkmark
& \checkmark/\checkmark
\\
NuTeV/CCFR $2\mu$ CC DIS
& (\checkmark/\checkmark)$^b$
& 
& \checkmark/-
& 
& 
\\
$pA$ DY
& \checkmark
& \checkmark
& \checkmark
& 
& \checkmark
\\
$\pi A$ DY
& 
& \checkmark
& 
&
\\
\midrule
\textcolor{blue}{\sc Collider data:}\ \\ 
  $Z$ bosons 
& \checkmark
& \checkmark
& \checkmark
& \checkmark
\\
  $W^\pm$ bosons
& \checkmark
& \checkmark
& \checkmark
& \checkmark
\\
Light hadrons 
& \checkmark
& \checkmark$^d$
& 
\\
-- Cut on $p_T$
& 3 GeV
& 3 GeV
\\
Jets
& 
& \checkmark
& \checkmark
\\
Prompt photons
& 
& 
& \checkmark
\\
Prompt D$^0$  
& \checkmark
& \checkmark
& \checkmark$^e$
\\
-- Cut on $p_T$
& 3 GeV
& 3 GeV
& 0 GeV
\\
Quarkonia ($J/\psi$, $\psi'$, $\Upsilon$)
& \checkmark
& 
& 
\\
\bottomrule
\end{tabular}
\begin{tabnote}
$^{\rm a}$ nCTEQ15HIX \cite{Segarra:2020gtj}; $^{\rm b}$ nCTEQ15$\nu$ \cite{Muzakka:2022wey}; $^{\rm c}$ through CT18A; $^{\rm d}$ only $\pi^0$ in DAu; $^{\rm e}$ only forward ($y>0$).
\end{tabnote}
\end{table}
\renewcommand{\arraystretch}{1}

\begin{figure}
\includegraphics[width=1.0\textwidth]{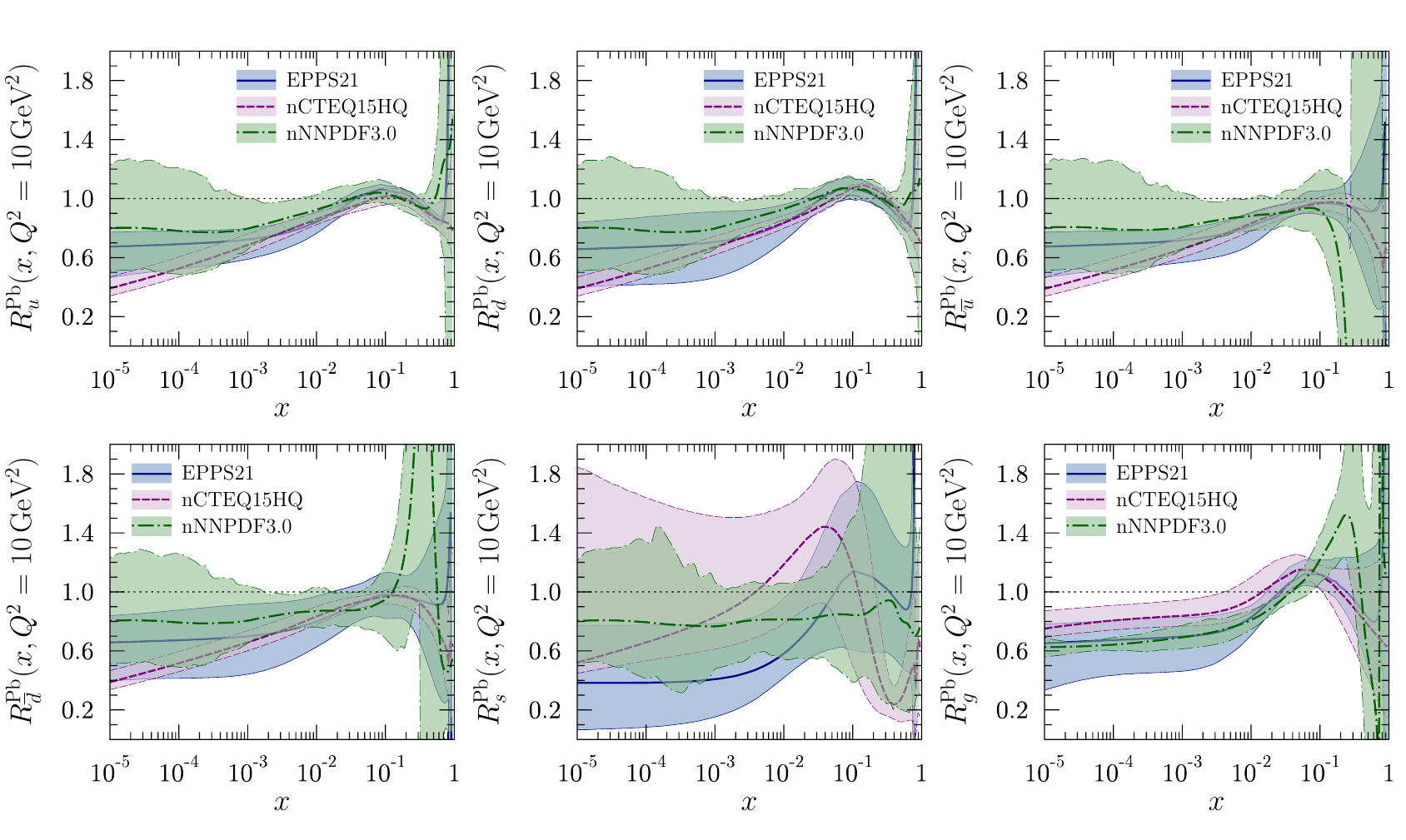}
\caption{Comparison of the $^{208}$Pb nuclear modifications resulting from the EPPS21 (full, blue) \cite{Eskola:2021nhw}, nCTEQ15HQ (dashed, red) \cite{Duwentaster:2022kpv} and nNNPDF3.0 (dot-dashed, green) \cite{AbdulKhalek:2022fyi} global analyses of nuclear PDFs, i.e.\ the PDFs of lead divided by the summed PDFs of 82 free protons and 126 free neutrons. Uncertainty bands correspond to 90\% CL.}
\label{fig:nPDFs}
\end{figure}

Figure \ref{fig:nPDFs} compares the nuclear modifications of the lead nucleus PDFs at $Q^2 = 10$ GeV$^2$ from EPPS21 (full, blue) \cite{Eskola:2021nhw}, nCTEQ15HQ (dashed, red) \cite{Duwentaster:2022kpv} and nNNPDF3.0 (dot-dashed, green) \cite{AbdulKhalek:2022fyi}. Qualitatively, there is good overall agreement between all three within the 90\% CL uncertainty bands (shaded areas). Closer inspection reveals nevertheless still significant differences both among the central values and the widths of the uncertainty bands in several distributions and $x$ regions. Due to the precise fixed-target NC DIS data, the nuclear modifications of $u$ and $d$ quarks (and to a lesser extent antiquarks) are well constrained at $x\sim 0.1$, though the widths of the error bands differ by at least a factor of two. Larger uncertainties remain in the Fermi motion region, in particular in nNNPDF3.0, where the $x$ dependence at the starting scale is not parametric. In the case of $\bar u$ and $\bar d$, the fit nCTEQ15HQ comes with the smallest uncertainties, which could, however, be due to not fitting $\bar u$ and $\bar d$ separately. Thanks to the LHC data, the gluon uncertainties are now much better constrained than in the previous rounds of global fits down to $x\sim10^{-5}$, and they also impact the sea (anti)quarks and their uncertainties at higher $Q^2$. Below $x\sim10^{-3}$, the nCTEQ15HQ and nNNPDF3.0 gluon uncertainty bands do not overlap, which will be reflected in some of the plots comparing theoretical predictions and LHC data in Sec.~\ref{sec:05}. The largest uncertainties are seen for the strange quark distributions, which are constrained only by -- to some extent problematic -- neutrino data and by LHC weak boson data, where the strange quark originates, however, mostly from gluon splittings. In \nameref{sect:Supplementarymaterial} we provide also a comparison of the absolute nuclear PDFs.

\section{Fixed-target data}
\label{sec:04}

We now turn to the detailed discussion of the available experimental data and their impact on global nuclear PDF analyses in roughly chronological order. We first focus on the early fixed-target NC and CC DIS and DY data, but review also the more recent JLab DIS data.

\subsection{Early DIS data and constraints on quarks}
\label{sec:4.1}

Measurements of fixed-target electron and muon NC DIS on various nuclei from SLAC, EMC and NMC, but also from the BCDMS, FNAL and HERMES experiments form the backbone of global nuclear PDF determinations \cite{Arneodo:1992wf}. Most of these data sets can be fitted with an excellent $\chi^2/$dof, but a few of them such as the 1988 EMC measurement of $F_2^{\rm Sn}/F_2^{\rm D}$ \cite{EuropeanMuon:1988lbf} are difficult to describe \cite{Hirai:2007sx,Duwentaster:2022kpv,AbdulKhalek:2022fyi,Helenius:2021tof,Khanpour:2020zyu,deFlorian:2011fp}. In contrast, the NMC $F_2^{\rm Sn}/F_2^{\rm C}$ data \cite{NewMuon:1996gam,NewMuon:1996yuf} can be fitted well \cite{Duwentaster:2022kpv,Eskola:2021nhw,AbdulKhalek:2022fyi,Helenius:2021tof,Khanpour:2020zyu}. Other outliers are e.g.\ the E665 data \cite{E665:1995xur} on $F_2^{\rm C}/F_2^{\rm D}$, $F_2^{\rm Ca}/F_2^{\rm D}$ and $F_2^{\rm Pb}/F_2^{\rm D}$, though e.g. the ratio $(F_2^{\rm Pb}/F_2^{\rm D})/(F_2^{\rm C}/F_2^{\rm D}) = F_2^{\rm Pb}/F_2^{\rm C}$ is consistent with the NMC data \cite{NewMuon:1996yuf}.

The kinematic reach $x \gtrsim 5\times 10^{-3}$ and $Q^2 \lesssim 140\,{\rm GeV}^2$ of these fixed-target data is naturally more limited than the one of the HERA experiments H1 and ZEUS, which form the bulk of the data in free proton analyses \cite{Kovarik:2019xvh,Ethier:2020way}. In addition, cuts are often applied in order to limit effects of target-mass and other higher-twist corrections, which could be larger in nuclear reactions \cite{Accardi:2004be}. At low $Q^2$, NC DIS is governed by virtual photon exchange. In the kinematic region of fixed-target experiments, the cross section
\begin{equation}
\frac{d^2\sigma^{lA}}{dxdQ^2} = \frac{4\pi\alpha^2}{Q^4} \left[F^A_2(x,Q^2) \left( \frac{y^2}{2} +  1-y-\frac{xyM^2}{s-M^2} \right) -xy^2 F^A_{\rm L}(x,Q^2) \right]
\end{equation}
is dominated by the structure function $F^A_2(x,Q^2)$, which in LO is sensitive only to the squared charge-weighted sum of quarks and antiquarks (cf.\ Sec.\ \ref{sec:01}). As the $lA$ DIS data that enter the global fits are given in terms of ratios 
\begin{equation}
\frac{d\sigma^{lA_1}}{d\sigma^{lA_2}} \approx \frac{F_2^{A_1}}{F_2^{A_2}} \,,
\end{equation}
where $A_2$ is typically D or carbon (C), the DIS data can mainly directly constrain the overall nuclear modification of valence and sea quarks. The contributions of gluons enter the cross section only at order $\alpha_s$, and the direct constraints for the gluon densities are therefore weak. However, the gluons drive the $Q^2$ dependence of $F_2^A(x,Q^2)$ at small values of $x$ \cite{Prytz:1993vr},
\begin{equation}
\frac{dF^A_2(x,Q^2)}{d\log Q^2} \approx \frac{10\alpha_s(Q^2)}{27\pi} xf^A_g(2x,Q^2) \,, \ x \rightarrow 0 \,.
\label{eq:35}
\end{equation}
Through this relation, it was understood early on that the $Q^2$ dependence of the ratios $F_2^{\rm Sn}/F_2^{\rm C}$ measured by NMC \cite{NewMuon:1996gam} around $x \approx 0.01\ldots 0.02$ and $Q^2 \approx 1 \ldots 10 \,{\rm GeV}^2$ can constrain the $A$ dependence of the gluon nuclear modifications \cite{Eskola:2002us}. In particular, a very strong $A$ dependence of gluons would contradict the measured positive $Q^2$ slopes of $F_2^{\rm Sn}/F_2^{\rm C}$. There are also similar HERMES data for $F_2^{\rm Kr}/F_2^{\rm D}$ \cite{HERMES:2002llm}, but the $Q^2$ lever arm is not as long in the perturbative regime. In principle, the longitudinal structure function $F^A_L$ carries a direct sensitivity to the gluon \cite{Armesto:2010tg}, but data are scarce \cite{HERMES:1999bwb}. Likewise, the cross sections for charm production would provide more direct information on the gluons -- modulo a possible intrinsic charm PDF \cite{Ball:2022qks} -- but not much data are available \cite{EuropeanMuon:1980wht,EuropeanMuon:1982xfn}.

Since $F_2^A$ probes predominantly the squared charge-weighted sum of quark PDFs, the flavor decomposition is also difficult to pin down. To understand this, let us write e.g. the valence quark distributions as
\begin{align}
f_{u_v}^{A} & = R^A_v \left( \frac{Z}{A} f_{u_v}^p + \frac{A-Z}{A} f_{d_v}^p\right) + 
\delta R^A_v \left( \frac{2Z}{A} - 1\right) \frac{f_{u_v}^p f_{d_v}^p}{f_{u_v}^p + f_{d_v}^p} \,, \label{eq:uvsd1} \\
f_{d_v}^{A} & = R^A_v \left( \frac{Z}{A} f_{d_v}^p + \frac{A-Z}{A} f_{u_v}^p\right) - 
\delta R^A_v \left( \frac{2Z}{A} - 1\right) \frac{f_{u_v}^p f_{d_v}^p}{f_{u_v}^p + f_{d_v}^p} \label{eq:uvsd2} \,, 
\end{align}
where $R^A_v \equiv ({R^{p/A}_{u_v} f_{u_v}^p + R^{p/A}_{d_v} d_{d_v}^p})/({f_{u_v}^p + f_{d_v}^p})$ is an average nuclear modification of the valence quarks and the difference is $\delta R^A_v \equiv R^{p/A}_{u_v} - R^{p/A}_{d_v}$. The first terms dominate and thus the large-$x$ data constrain very tightly the average modification $R^A_v$. Having data for several different combinations of $Z$ and $A$, combined with the fact that the nuclear effects are expected to scale with $A$, will give also constraints on $\delta R^A_v$ and thereby to the mutual differences of nuclear effects in up and down valence quarks. The same reasoning naturally applies for the up and down sea quarks. Eqs.~\ref{eq:uvsd1} and \ref{eq:uvsd2} also clearly demonstrate the anticorrelation of the nuclear effects between up and down quarks. For an isoscalar nucleus $Z=A/2$, the up and down quark distributions are always equal.

To facilitate the interpretation of nuclear effects, many early NC DIS experiments corrected their data for isospin effects, i.e.\ for the unequal numbers of protons and neutrons in heavy nuclei compared to the deuteron. These data were long taken at face value -- and are still done so e.g. in nNNPDF3.0 -- and fitted by setting $Z=N=A/2$. However, this is not necessary in global fits and has in the past even caused some confusion about the nuclear valence quark modification \cite{Eskola:2016oht}. Comparing $F_2^A = [ZF_2^{p/A}+(A-Z)F_2^{n/A}]/A$ (cf.\ Eq.\ \ref{eq:11}) with the isoscalar expression $\hat{F}_2^A = [F_2^{p/A}+F_2^{n/A}]/2$ leads to
$\hat{F}_2^A  =  \beta F_2^{A}$ with
\begin{equation}
\beta  = \frac{A}{2} \left(1+\frac{F_2^{n/A}}{F_2^{p/A}}\right) / \left(Z + (A-Z)\frac{F_2^{n/A}}{F_2^{p/A}}\right) \, .
\end{equation}
The experiments then assumed $F_2^{n/A}/F_2^{p/A} = F_2^n/F_2^p$ and parameterized this ratio from DIS data on protons and deuterons with, e.g., $1-0.8x$ \cite{Gomez:1993ri} or $0.92-0.86x$ \cite{EuropeanMuon:1992pyr}. These functions then allow to calculate $\beta$ and either apply it also to the theoretical calculations \cite{Walt:2019slu} or to remove the isoscalar correction from the data and fit the true nuclei \cite{Segarra:2020gtj,Eskola:2016oht}.

\subsection{DIS at high $x$ and nuclear effects in the deuteron}
\label{sec:4.2}

Recent precise JLab measurements taken with $6 - 10$~GeV electrons on various nuclear targets \cite{Seely:2009gt,Arrington:2021vuu,HallC:2022rzv,CLAS:2019vsb} at low to intermediate $Q^2$ and $W^2$ can considerably reduce the nuclear PDF uncertainties in the high $x$ region \cite{Segarra:2020gtj,Paukkunen:2020rnb}. This region requires, however, a good control over potential target-mass (cf.\ Sec.\ \ref{sec:2.2}) and other higher-twist corrections (cf.\ Sec.\ \ref{sec:2.3}), hadronic resonances and (similarly to the other fixed-target DIS data) nuclear effects in the deuteron. Target-mass corrections scale with powers of $M_N^2/Q^2$ and are thus suppressed even for heavy nuclei. Their leading effect is a shift in the probed momentum fraction $x_N$ to the Nachtmann variable $\xi_N$ \cite{Ruiz:2023ozv}. When the kinematic cuts are relaxed from $Q^2>4$ GeV$^2$, $W^2>12.25$ GeV$^2$ to $Q^2>1.69$ GeV$^2$, $W^2>2.89$ GeV$^2$, the subleading target mass effects provide a uniform shift of below one percent for all nuclei, leaving the ratios $F_2^A/F_2^{\rm D}$ unaffected \cite{Segarra:2020gtj}. Some global analyses therefore include the JLab data with lower kinematic cuts \cite{Segarra:2020gtj,Eskola:2021nhw,Khanpour:2020zyu,Paukkunen:2020rnb}. The compatibility of the JLab data with global fits can be taken as evidence that other higher-twist effects can be neglected \cite{Eskola:2021nhw}. They can, however, also be parameterized as
\begin{align}
F_2^A(x,Q) & \to  F_2^{A}(x,Q) \left[ 1+ \frac{A^{1/3} h_0 x^{h_1} (1+ h_2 x)}{Q^2} \right] \,, \label{eq:ht} 
\end{align}
where the values $\{h_0, h_1, h_2\}= \{-3.3~{\rm GeV}^2, 1.9, -2.1\}$ come from the CJ15 proton PDF analysis \cite{Accardi:2016qay} and $A^{1/3}$ scaling is assumed (cf.\ Sec.\ \ref{sec:2.3}). This form of higher-twist corrections leads to a slight reduction of $F^A_2$ at intermediate $x\,{\sim}\, 0.3$ and a substantial enhancement at high $x$ and $Q^2<16 \, {\rm GeV}^2$, and it improves the global $\chi^2$ by around 3\% within the nCTEQ15HIX global analysis  \cite{Segarra:2020gtj}.

In the resonance region at very low $W^2\in[1.21;2.89]$ GeV$^2$, the nuclear effects at large $x$ are surprisingly similar to those in DIS, which may signal the applicability of quark hadron duality due to the averaging over nuclear resonances \cite{Arrington:2003nt}. As described in the introduction, Fermi motion can be accounted for by a convolution of the nucleon structure function with the nucleon momentum distribution or effectively by a rescaling of the variable $x$. The rise of $F_2^A/F_2^{\rm D}$ at large $x$ can be well described by the parameterization \cite{Segarra:2020gtj}
\begin{equation}
 x' =  x - \varepsilon \, x^\kappa \log_{10}A \ .
 \label{eq:rescale}
\end{equation}
As the PDFs decrease with $x$, the negative shift ensures that the transformed function is larger than the unmodified one and non-vanishing as $x \to 1$. The overall size of the rescaling effect is controlled by $\varepsilon$, $\kappa>0$ ensures that only the large-$x$ region is modified, and the $\log_{10} A$ term implies an increasing modification across the full range of nuclear $A$ values from the proton $(A=1)$ to lead $(A=208)$. A good description of the JLab data is obtained with $\kappa=10$ and $\epsilon \sim 0.03$. While this suggests that it may be possible to expand the kinematic reach to $W^2<2.89$ GeV$^2$, the resonance region is currently avoided in all global fits (cf.\ Tab.\ \ref{tab:01}). 

Given that the nuclear DIS data are usually presented as ratios $F_2^A/F_2^{\rm D}$ and that most global fits of proton PDFs use deuteron data as well, a good control over the nuclear effects in the deuteron is required. The deuteron is much more loosely bound than heavier nuclei and therefore often approximated as an isoscalar combination of a free proton and neutron. However, its structure at large $x$ is still modified by Fermi motion, nuclear binding and off-shell effects, while at small $x$ rescattering still induces some shadowing. These nuclear effects are of the order of a few percent in the available DIS data and below 1\% in the available DY data. They can be accounted for in different ways: one option is to use a free-proton baseline that fitted deuteron DIS data without nuclear corrections \cite{Eskola:2021nhw}. In this case the deuteron nuclear effects are, to some extent, fitted into the $u$ vs.\ $d$ quark flavor separation. Then no additional nuclear correction should be applied, although also this has been done, and in this case the effect of double counting should be quantified \cite{Khanpour:2020zyu}. Alternatively, one can rescale the fitted $F_2^A/F_2^{\rm D}$ data by a ratio $F_2^{\rm D}/F_2^p$ \cite{Segarra:2020gtj} as modeled, e.g., within the CJ15 global proton analysis \cite{Accardi:2016qay} (cf.\ also \cite{Martin:2012da}). If this is done with a free proton PDF that already includes deuteron data \cite{Segarra:2020gtj}, the same deuteron correction should be applied in both cases \cite{Owens:2007kp}. A third possibility is to fit the free proton without deuteron data and then the deuteron in the same way as the other nuclei \cite{Helenius:2021tof}. The theoretical treatment of the deuteron affects the description of all NC DIS data and therefore also the question of the compatibility of CC DIS data with NC DIS and electroweak boson production at the LHC (cf.\ Secs.\ \ref{sec:4.4} and \ref{sec:5.1}).

\subsection{Drell-Yan process and constraints on antiquarks}
\label{sec:4.3}

The Drell-Yan process, i.e.\ the inclusive production of electroweak gauge bosons in hadron collisions, followed by a leptonic decay of the gauge boson, has been of enormous historical importance for the quark flavor separation in protons \cite{Gao:2017yyd}. For heavier nuclei, fixed-target measurements have been made by the FNAL E605 \cite{Moreno:1990sf}, E772 \cite{Alde:1990im} and E866 \cite{NuSea:1999egr} 
experiments in $pA$ collisions covering several nuclei from carbon to tungsten in the kinematic range $x>10^{-2}$ and dilepton mass $M_{ll}<15$~GeV. At such low values of $M_{ll}$, well below the $Z$-boson peak, the DY process is dominated by an off-shell intermediate photon with the cross section differential in the lepton pair rapidity $y_{ll}$ and $M_{ll}$ given by
\bea 
\label{eq:dy1}
\frac{d^2\sigma^{pA}_{\rm DY}}{dM_{ll}dy_{ll}}\sim \sum_q
e_q^2\le f^{p}_q(x_1)f^A_{\bar{q}}(x_2)+f^{p}_{\bar{q}}(x_1) f^A_q(x_2) \re &{\rm with}&
x_{1,2} = \frac{M_{ll}e^{\pm y_{ll}}}{\sqrt{s}} \,, 
\eea
which tests a squared charge-weighted combination of quarks and antiquarks. Fixed-target experiments generally have larger acceptance in the $x_1\gg x_2$ region, where $x_1$ is defined with respect to the proton beam, and in this case the first term in Eq.~\ref{eq:dy1} is the dominant one, where $f^{p}_q(x_1)$ is mostly determined by the valence quark content of the proton. For isoscalar nuclei, $f^A_{\bar{u}} = f^A_{\bar{d}}$, and the cross section ratios between $pA$ and $p$D collisions, measured in the E772 and E866 experiments, become
\begin{equation}
\label{eq:dy}
\frac{d\sigma_{\rm DY}^{pA}}{d\sigma_{\rm DY}^{p{\rm D}}}_{\Big| {\rm isoscalar} \ A} \approx \frac{f^A_{\bar{u}}(x_2)}{f^{\rm D}_{\bar{u}}(x_2)} = \frac{f^A_{\bar{d}}(x_2)}{f^{\rm D}_{\bar{d}}(x_2)} \,. 
\end{equation}
As a result, the measured DY ratios are sensitive to the nuclear modifications of sea-quark distributions at $x_2\sim 0.03...0.3$ \cite{Ellis:1990ti,Martin:1998sq}. In principle, the dependence on $M_{ll}$ should also retain sensitivity to the gluon through the DGLAP evolution. Combining the photon-mediated DIS and DY measurements gave historically the first chance to disentangle the nuclear effects in valence and sea quarks leading to the conclusion that there was not such a clear antishadowing effect in sea quarks as there was for valence quarks. The E605 data \cite{Moreno:1990sf} for $p$Cu collisions is given in terms of absolute cross sections and is often used in fits of proton PDFs. In the future, fixed-target $pA$ data from the FNAL E906/SeaQuest experiment \cite{Lin:2017eoc} are expected to improve the precision of the available data. The renewed facilities at RHIC should also be able to provide new measurements on the DY process \cite{Helenius:2019lop}, and similar measurements are planned at the LHCb experiment at the LHC \cite{LHCb:2018qbx}.

In principle, the pion-nucleus DY process has the potential to constrain the flavor decomposition of the valence quarks \cite{Paakkinen:2016wxk}. It depends on the pion PDFs, but this dependence cancels largely in ratios of nuclear cross sections. Unfortunately it turns out that the precision of the $\pi A$ DY data from the CERN NA3 \cite{NA3:1981yaj}, NA10 \cite{NA10:1987hho} and FNAL E615 \cite{Heinrich:1989cp} experiments is not high enough to provide significant discrimination power on top of the DIS data. However, the new CERN-based facility AMBER \cite{Adams:2018pwt} may be able to improve upon the current precision.

\subsection{Neutrino DIS data and flavor separation}
\label{sec:4.4}

Due to the weak nature of neutrino interactions, heavy nuclear targets such as iron or lead have traditionally been used to obtain CC DIS data with sufficient statistics. Despite the nuclear targets, these data have routinely been included in global analyses of proton PDFs, with or without nuclear corrections \cite{Ball:2018twp}, and have formed the principal constraint for a possible $s$ vs.\ $\bar s$ asymmetry \cite{NuTeV:2007uwm}. In addition, determinations of the weak mixing angle in neutrino DIS \cite{NuTeV:2001whx} have relied on a sufficient understanding of the nuclear structure \cite{Hirai:2004ba,Eskola:2006ux,ParticleDataGroup:2022pth}. In the case of CC neutrino DIS, the differential cross section is
\beq
 \frac{d^2\sigma^{\nu,\bar\nu A}}{dxdy}\!=\!\frac{G_F^2M_W^4}{2\pi xyQ^2} \lr\!\frac{Q^2}{Q^2\!+\!M_W^2}\!\rr^2\!\le\!\lr1\!-\!y\!-\!\frac{x^2y^2 M^2_N}{Q^2}\rr\! F^{\nu,\bar \nu A}_2\!\!+\!y^2xF^{\nu,\bar \nu A}_1\pm\!\lr y\!-\!\frac{y^2}{2}\rr xF^{\nu,\bar \nu A}_3\re\!,
\eeq
where $G_F$ is the Fermi constant, $M_W$ the $W$-boson mass and the $+$ ($-$) sign is taken for incoming (anti-)neutrinos. At LO and high $Q^2$, 
\begin{align}
d^2\sigma^{\nu A} & \propto \left( f_d^A + f_s^A + f_b^A \right) + (1-y)^2 \left( f_{\bar u}^A + f_{\bar c}^A \right) \,, \\
d^2\sigma^{\bar\nu A} & \propto \left( f_{\bar d}^A + f_{\bar s}^A + f_{\bar b}^A \right) + (1-y)^2 \left( f_{u}^A + f_{c}^A \right) \,.
\end{align}
Due to the suppressing factor $(1-y)^2$, there is an increased sensitivity to the strange quark distribution in comparison to NC charged-lepton DIS, particularly for antineutrinos. In addition, the up and down quarks enter the cross sections with different weights than in the case of NC charged-lepton DIS. Adding the neutrino data thus helps in constraining the differences between nuclear effects in up and down quarks (cf.\ Eqs.~\ref{eq:uvsd1} and \ref{eq:uvsd2}). However, in a global fit these data are also sensitive to the assumed proton PDFs.

Data on inclusive neutrino DIS have been taken by e.g. by CDHSW \cite{Berge:1989hr}, CCFR \cite{CCFRNuTeV:2000qwc} and NuTeV \cite{NuTeV:2005wsg} on iron and by CHORUS on lead \cite{CHORUS:2005cpn}. Also the charm production has been measured through muonic decays of produced charmed hadrons (cf.\ Fig.\ \ref{fig:01}) \cite{NuTeV:2001dfo,NOMAD:2013hbk}. In principle, all of these data should be relevant for nuclear PDFs, but are only partially included in global analyses due to concerns about possible mutual tensions between the neutrino data sets, tensions with the charged-lepton DIS data, and due to the fact that part of these neutrino data are in some cases already used in the proton PDF fits that are used as baselines in the fits of nuclear PDFs. One of the difficulties is also that there are no references from $\nu p$ or $\nu$D scattering, so that the data are reported as absolute cross sections.

\begin{figure}
\centering
\includegraphics[width=\textwidth]{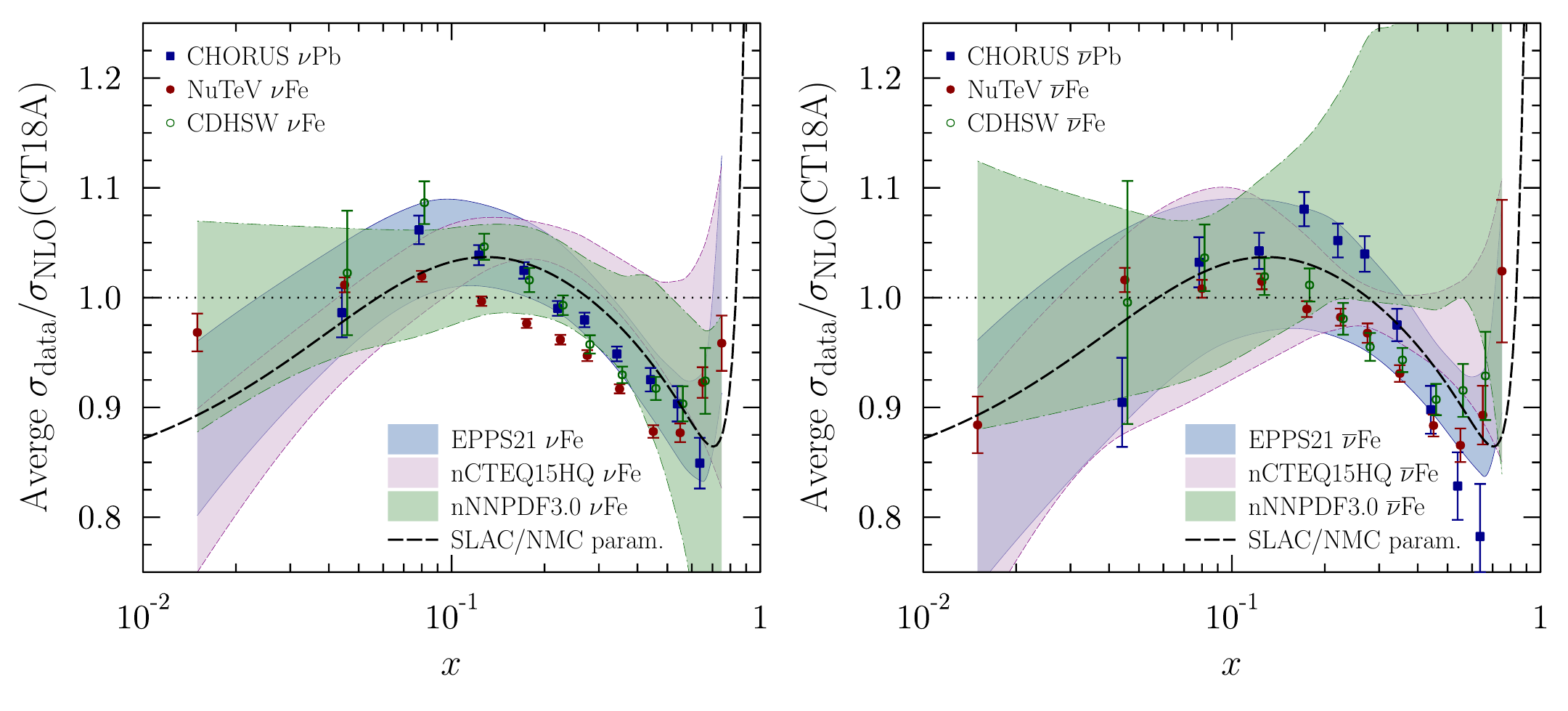}
\caption{Average ratios of neutrino (left) and antineutrino (right) cross sections as measured by CHORUS \cite{CHORUS:2005cpn}, NuTeV \cite{NuTeV:2005wsg} and CDHSW \cite{Berge:1989hr} to a theoretical NLO prediction with CT18A PDFs within the kinematic range $Q^2>4$ GeV$^2$ and $W^2>12.25 \,{\rm GeV}^2$. The data are compared with EPPS21 \cite{Eskola:2021nhw}, nCTEQ15HQ \cite{Muzakka:2022wey} and nNNPDF3.0 \cite{AbdulKhalek:2022fyi} predictions. The SLAC/NMC NC DIS parameterization in Eq.~\ref{eq:02} is also shown as a reference.}
\label{fig:05}
\end{figure}

Figure \ref{fig:05} shows neutrino (left) and antineutrino (right) cross sections divided by the theoretical NLO predictions with CT18A proton PDFs in the SACOT-$\chi$ scheme,  including approximate target mass and electroweak corrections as used in Ref.\ \cite{Paukkunen:2010hb}. The ratios are evaluated as weighted averages over $Q^2>4 \, {\rm GeV}^2$ and $W^2>12.25 \, {\rm GeV}^2$ as in Ref.\ \cite{Muzakka:2022wey}. The obtained ratios are compared with predictions for NuTeV data using EPPS21 \cite{Eskola:2021nhw}, nCTEQ15HQ \cite{Muzakka:2022wey} and nNNPDF3.0 \cite{AbdulKhalek:2022fyi}. Also the SLAC/NMC NC DIS parameterization (Eq.~\ref{eq:02}) is shown for comparison. Tensions between different data sets, nuclear PDFs, and data and theory are clearly visible. The largest differences between nuclear PDFs occur in the case of $\bar \nu$ DIS at $x \gtrsim 0.2$, where the nNNPDF3.0 values are significantly above those of EPPS21 or nCTEQ15HQ. This can be explained by the large enhancement of $\bar d$ and $s$ densities of nNNPDF3.0 in comparison to EPPS21 or nCTEQ15HQ (cf.~\nameref{sect:Supplementarymaterial}). From the data sets, in particular the NuTeV neutrino data stand out from the others, but also larger deviations between the CHORUS and CDHSW data have been observed, if less restrictive kinematic cuts are imposed and electroweak corrections are neglected \cite{Muzakka:2022wey}. To some degree the observed tensions can be alleviated by normalizing the data by the cross sections integrated over $x$ and $y$ \cite{Paukkunen:2010hb,Paukkunen:2013grz}, by neglecting the NuTeV systematic error correlations, or by introducing additional cuts in $x$ \cite{Muzakka:2022wey}. The current consensus seems to be that at least the CHORUS data can be included in global analyses without significant tensions. In addition, the charm dimuon data are used in nCTEQ15$\nu$ \cite{Muzakka:2022wey} and nNNPDF3.0 \cite{AbdulKhalek:2022fyi} and the CDHSW data in TUJU21 \cite{Helenius:2021tof} and KSASG20 \cite{Khanpour:2020zyu}. There have been speculations about differences in nuclear shadowing in CC and NC processes \cite{Kopeliovich:2012kw}, at least at low $Q^2$ \cite{Frankfurt:2011cs}, but $W^\pm$ and $Z$ production at the LHC probing nuclear PDFs at significantly higher $Q^2$ can be fitted well in global analyses.

In the future, novel neutrino-nucleus DIS data may become available through dedicated experiments measuring neutrinos produced in high-luminosity $pp$ collisions at the LHC. Indeed, the first observations of such collider neutrinos have already been made by the FASER \cite{FASER:2023zcr} and SND@LHC \cite{SNDLHC:2023pun} collaborations. The impact of such future measurements on nuclear PDFs has been considered recently in Ref.~\cite{Cruz-Martinez:2023sdv}.

\section{Collider data}
\label{sec:05}

We now turn to the discussion of the LHC ($p$Pb), but also RHIC (DAu) measurements used in global fits of nuclear PDFs. Within collinear factorization, the $p$A (or DA) cross sections
\begin{equation}
d\sigma(pA \rightarrow \mathcal{O} + X) = \sum_{\rm i,j[,k]} f_i^{p} \otimes f_j^{A} \otimes d\hat\sigma(ij \rightarrow \mathcal{O},[k] + X) \ [\,\otimes \, D_k^\mathcal{O}\,]
\end{equation}
for the observable $\mathcal{O}$ involve convolutions of (nuclear) PDFs $f_{i,j}^{p,A}$ with perturbative partonic cross sections $d\hat\sigma$ and in the case of inclusive light or heavy-flavored hadron ($h$) production non-perturbative fragmentation functions (FFs) $D_k^h$. These are numerically costly and must in many cases be evaluated with precomputed grids \cite{Carli:2010rw}. Since one of the colliding objects is a proton and also the nuclear PDFs depend on the proton, absolute LHC cross sections depend directly and indirectly on the proton PDFs. To reduce this dependence and also cancel other theoretical and experimental uncertainties, the nuclear modification ratio
\begin{align}
R_{pA} & \equiv d\sigma(pA \rightarrow \mathcal{O} + X) \big / d\sigma(pp \rightarrow \mathcal{O} + X)
\end{align}
and forward-to-backward ratio
\begin{align}
R_{\rm FB} & \equiv d\sigma(pA \rightarrow \mathcal{O} + X)_{\big|y>0} \, \big / d\sigma(pA \rightarrow \mathcal{O} + X)_{\big|y<0}
\end{align}
are often introduced. Here $y$ refers to the rapidity of the observable $\mathcal{O}$. 

In fits of collider data, normalization uncertainties originating from the luminosity measurements play a special role. For example, the measured and calculated $R_{p{\rm Pb}}$ for hadron production in the $y \gg 0$ region (small $x_N$) at the LHC is often rather flat, and changes in the nuclear PDFs can be compensated by treating the normalization uncertainty as a correlated systematic uncertainty, see e.g. Ref.~\cite{Muzakka:2022wey}. If the luminosity uncertainty is common for $y > 0$ and $y < 0$, the $R_{\rm FB}$ is free from this additional freedom. Today, all global fits of nuclear PDFs account for the systematic normalization uncertainties. 

\subsection{Electroweak bosons}
\label{sec:5.1}

\renewcommand{\arraystretch}{1.05}
\begin{table}
\centering
\tabcolsep4.0pt
\caption{Summary of  $Z$, $W^\pm$, and low invariant mass $Z/\gamma^*$ rapidity distributions available from LHC $p$Pb collisions.}
\label{tab:02}
\hspace{-1cm}\begin{tabular}{lcccccc}
\\
\toprule
{\sc Data set}
& nCTEQ15HQ~\cite{Duwentaster:2022kpv}
& EPPS21~\cite{Eskola:2021nhw} 
& nNNPDF3.0~\cite{AbdulKhalek:2022fyi}
& TUJU21~\cite{Helenius:2021tof}
& KP16~\cite{Ru:2016wfx}
\\
\midrule
\textcolor{blue}{\sc Run-I:}\\
ATLAS $Z$ \cite{ATLAS:2015mwq} &\checkmark         &\checkmark       &\checkmark        &\checkmark &\checkmark\\
CMS $Z$ \cite{CMS:2015zlj}  &\checkmark         &\checkmark       &\checkmark        &\checkmark &\checkmark\\
ALICE $Z$ \cite{ALICE:2016rzo}  &          &        &\checkmark$^{\rm b}$ \\
LHCb $Z$ \cite{LHCb:2014jgh}  &\checkmark&        &\checkmark$^{\rm b}$ \\
\midrule
ATLAS $W^\pm$ \cite{TheATLAScollaboration:2015lnm}$^{\rm c}$  &\checkmark &&& &\checkmark\\
CMS $W^\pm$ \cite{CMS:2015ehw}  &\checkmark         &\checkmark       &\checkmark & \\
ALICE $W^\pm$ \cite{ALICE:2016rzo}  &\checkmark&        &\checkmark$^{\rm b}$ \\
\midrule
\textcolor{blue}{\sc Run-II:}\\
CMS $Z$ \cite{CMS:2021ynu}  &          &        &\checkmark$^{\rm b}$ \\
CMS $Z/\gamma^*$ \cite{CMS:2021ynu} &          &        &\checkmark$^{\rm b}$ \\
ALICE $Z$ \cite{ALICE:2020jff}  &          &        &\checkmark$^{\rm b}$ \\
LHCb $Z$ \cite{LHCb:2022kph}  & \\
\midrule
CMS $W^\pm$ \cite{CMS:2016qqr,CMS:2019leu}  &\checkmark         &\checkmark$^{\rm a}$       &\checkmark        &\checkmark \\
ALICE $W^\pm$ \cite{ALICE:2022cxs}  & \\
\bottomrule
\end{tabular}
\begin{tabnote}
$^{\rm a}$ added in EPPS21 \cite{Eskola:2021nhw}; $^{\rm b}$ added in nNNPDF3.0 \cite{AbdulKhalek:2022fyi}; $^{\rm c}$ unpublished.
\end{tabnote}
\end{table}
\renewcommand{\arraystretch}{1}

The first global analysis of nuclear PDFs to include $W$ and $Z$ boson data from $p$Pb collisions was EPPS16 \cite{Eskola:2016oht}. However, the impact of the Run-I ATLAS \cite{ATLAS:2015mwq} and CMS \cite{CMS:2015zlj,CMS:2015ehw} data was still rather limited due to low statistics. Thereafter, measurements for these electroweak processes have been published by all four LHC experiments at $\sqrt{s}=5.02$ TeV (Run-I) and by ALICE, CMS and LHCb at $\sqrt{s}=8.16$ TeV (Run-II). They are now, in different combinations, used in all recent global analyses \cite{Duwentaster:2022kpv,Eskola:2021nhw,AbdulKhalek:2022fyi} and also in the NNLO and model-dependent fits TUJU21 \cite{Helenius:2021tof} and KP16 \cite{Ru:2016wfx}. The available data are summarized in Tab.\ \ref{tab:02}. 

\begin{figure}
\includegraphics[width=\linewidth]{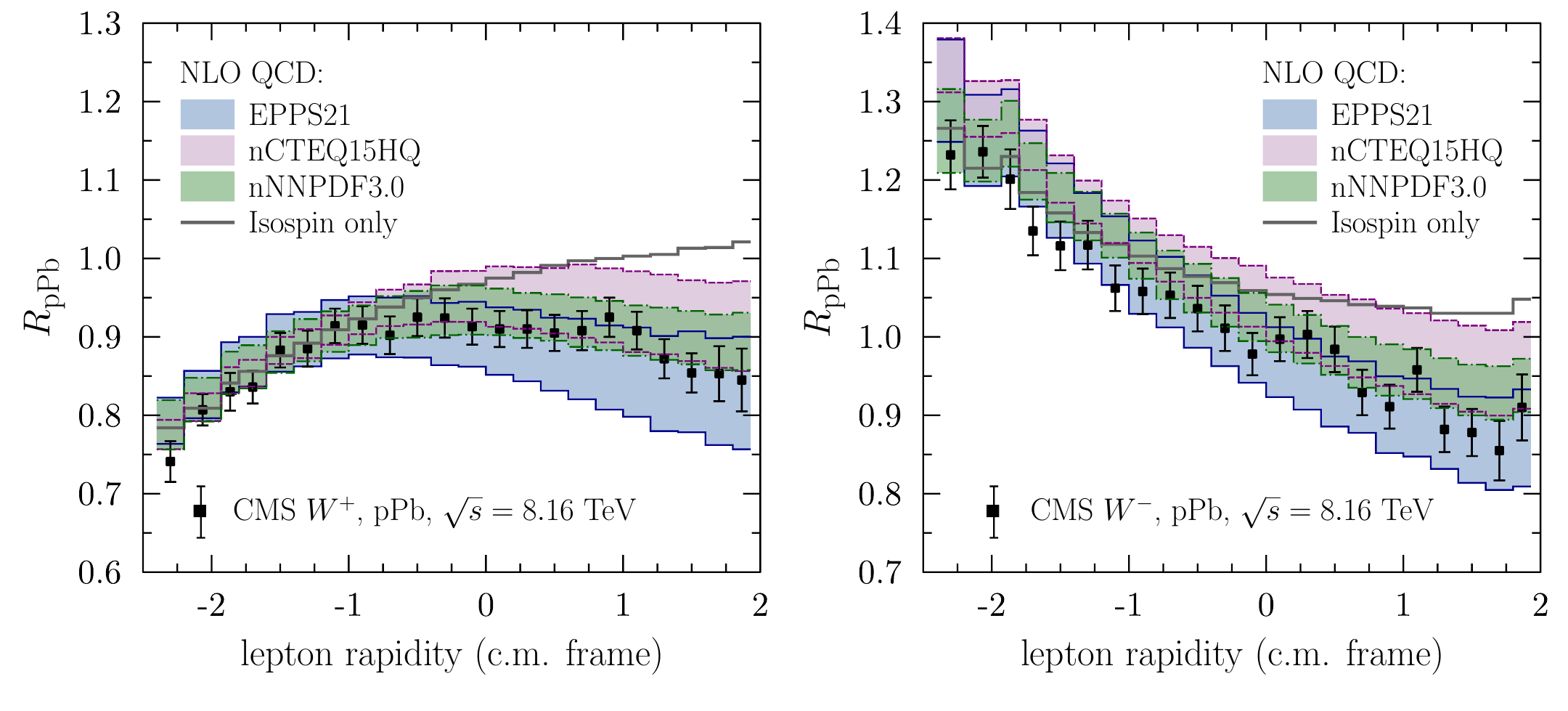}
\caption{Nuclear modification ratios for $W^+$ (left) and $W^-$ bosons (right) at CMS Run-II \cite{CMS:2016qqr,CMS:2019leu} compared with EPPS21 \cite{Eskola:2021nhw}, nCTEQ15HQ \cite{Duwentaster:2022kpv}, nNNPDF3.0 \cite{AbdulKhalek:2022fyi} and a calculation with 82 free protons and 126 free neutrons.}
\label{figW}
\end{figure}

At the moment, the most stringent constraints come from the Run-II CMS $W$ data \cite{CMS:2019leu}. Fig.~\ref{figW} compares the nuclear modification ratio $R_{p{\rm Pb}}$ constructed from the CMS Run-II \cite{CMS:2016qqr,CMS:2019leu} measurements with NLO calculations using the EPPS21 \cite{Eskola:2021nhw}, nCTEQ15HQ \cite{Duwentaster:2022kpv} and nNNPDF3.0 \cite{AbdulKhalek:2022fyi} nuclear PDFs. While the EPPS21 analysis included the shown $R_{p{\rm Pb}}$ data, the nCTEQ15HQ and nNNPDF3.0 analyses fitted absolute $p$Pb cross sections. As one can see, the spread between the different predictions is still rather significant. In comparison to a calculation with no nuclear effects, i.e.\ 82 free protons and 126 free neutrons, the data indicates a clear sign of shadowing at forward rapidities or $x \ll 1$. The relative ordering of EPPS21, nCTEQ15HQ and nNNPDF3.0 values follows the one of the corresponding gluon shadowing in Fig.~\ref{fig:nPDFs}. Also, as can be seen from Fig.~\ref{fig:nPDFs}, even after inclusion of these electroweak data, the overall variation in the strange quark PDF is still quite significant, which indicates that the constraints for the nuclear strange quark PDFs are still not very strong. 

The CMS Run-II measurement for $Z$ boson production \cite{CMS:2021ynu} reports similarly small uncertainties to the $W$ boson measurement. However, it is not possible to obtain a good quantitative description of these data with any nuclear PDFs due to large fluctuations of the data around midrapidity ($y_{ll}=0$), which lead e.g.\ to a $R_{\rm FB}$ that does not tend to unity towards $y_{ll} \rightarrow 0$ as one would expect. Along with the on-shell $Z$ production, CMS measured also low-mass cross sections in the window $15\,{\rm GeV} < M_{ll} < 60$ GeV. Within the TUJU21 analysis \cite{Helenius:2021tof} it was noticed that to simultaneously reproduce the normalization of the CMS low-mass and $Z$ cross sections, the NNLO QCD corrections appear to be necessary. This is the first time the necessity of NNLO corrections is seen in the case of $p$Pb collisions. 

Dielectron pairs have also been measured by ALICE at Run-I in the low-mass region $M_{ll}<3$ GeV and with $0<p_{T,ll}<8$ GeV \cite{ALICE:2020mfy}, which is in principle very sensitive to the gluon density and avoids the fragmentation contribution present for real photons \cite{Berger:1998ev,Klasen:2013ulb,Brandt:2014vva}. Currently the data are unfortunately still dominated by the heavy-flavor ($c$, $b$) decay background, but the statistics should be improved in Run-II and the background reducible with heavy-flavor tagging, in particular in LHCb \cite{LHCb:2018qbx}.

\subsection{Photons}
\label{sec:5.2}

Another electroweak probe of nuclear PDFs is the prompt production of real photons with finite transverse momentum $p_T$ \cite{Arleo:2007js,BrennerMariotto:2008st,Arleo:2011gc,Helenius:2014qla}. It proceeds directly through quark-antiquark annihilation ($q\bar{q}\to g\gamma$) and the QCD Compton process ($qg\to q\gamma$), which dominates at large $p_T$ \cite{Klasen:2002xb}. The radiation of massless photons from quarks in pure QCD processes gives rise to a photon fragmentation contribution \cite{Gluck:1992zx,Bourhis:1997yu,Klasen:2014xfa}, which is important at small $p_T$. Isolating the photon and thus reducing the surrounding hadronic energy (e.g.\ in a cone) suppresses the fragmentation component and also the non-prompt photon background from hadronic (in particular pion) decays \cite{Frixione:1998jh}.

Direct photon production in proton (and pion) nucleus collisions was first measured in fixed-target mode at the Fermilab E706 experiment \cite{FermilabE706:2004emk}. Also PHENIX \cite{PHENIX:2012krx} and STAR \cite{STAR:2009qzv} measurements in DAu collisions at RHIC are available. At the LHC, high-$p_T$ isolated photons in $p$Pb \cite{ATLAS:2019ery} collisions have been measured by the ATLAS collaboration. Ratios of these ATLAS cross section measurements to the corresponding $pp$ data \cite{ATLAS:2016fta} have a reduced sensitivity to missing higher-order effects, fragmentation functions and proton PDFs. They can be reasonably described at NLO QCD and are therefore included in the nNNPDF3.0 \cite{AbdulKhalek:2022fyi} analysis. The absolute $pp$ and $p$Pb cross sections are, however, underestimated by NLO QCD by up to 30\% at the lowest values of $p_{\rm T} \sim 20\,{\rm GeV}$. This could indicate the necessity to include NNLO corrections \cite{Campbell:2016lzl}. The impact of the ATLAS prompt photon data in the global fit is small in comparison to dijet and heavy flavor production due to larger uncertainties. Looking into the future, the ALICE collaboration has proposed to build a new forward calorimeter (FoCal) for LHC Run-IV \cite{ALICE:2020mso,ALICE:2023fov}, optimized for direct photons in the rapidity region $3.2 < \eta < 5.8$.

\subsection{Light hadrons}
\label{sec:5.3}

As discussed in Sec.~\ref{sec:04}, before the beginning of the LHC era not much was known about the nuclear gluon PDFs. The first direct evidence for the presence of shadowing, antishadowing and the EMC effect in gluons came from inclusive hadron production in D$A$ collisions at RHIC. This process involves a gluon contribution already at LO and is therefore a candidate to constrain the nuclear gluons in the perturbative region, i.e.\ when the transverse momentum $p_T$ of the hadron is sufficiently large. This possibility was first discussed in Refs.\ \cite{Eskola:2007my,deFlorian:2003qf,Vogt:2004hf} in the light of early RHIC data \cite{BRAHMS:2003sns,BRAHMS:2004xry,PHENIX:2003qdw,STAR:2003pjh}. The first global analysis to fit this type of data was EPS08 \cite{Eskola:2008ca}, which included negatively-charged hadron BRAHMS \cite{BRAHMS:2004xry} as well as PHENIX \cite{PHENIX:2003qdw,PHENIX:2006mhb} and STAR \cite{STAR:2006xud} pion data. It was, however, noticed that the rapidity dependence of the BRAHMS data at low values of $p_T \gtrsim 2 \, {\rm GeV}$ was too strong to be optimally reproduced within a global fit, inducing tensions with the NMC data for $F_2^{\rm Sn}/F_2^{\rm C}$ \cite{NewMuon:1996gam}.  These negatively-charged hadron BRAHMS data were eventually dropped from the EPS09 analysis \cite{Eskola:2009uj}, when it was noticed that it was difficult to reproduce even the $pp$ reference data. The later EP(P)S analyses \cite{Eskola:2021nhw,Eskola:2016oht} have retained only PHENIX $\pi^0$ data \cite{PHENIX:2006mhb}, while nCTEQ15 \cite{Kovarik:2015cma} included these and updated STAR $\pi^0$ data \cite{STAR:2009qzv}.  

Inclusive hadron production is not only sensitive PDFs, but also to the final-state hadronization encoded in the parton-to-hadron FFs. In the EP(P)S and nCTEQ fits, the FFs are taken from global fits of hadron production in $e^+e^-$, $eN$, and $pp$ collisions \cite{Kniehl:2000fe,Aidala:2010bn,deFlorian:2014xna,deFlorian:2017lwf,Moffat:2021dji}. The sensitivity to FFs was recently studied within the nCTEQ15WZ+SIH \cite{Duwentaster:2021ioo} analysis, which also propagated the FF uncertainties, when available, into the fit. Combining the latest RHIC DAu data on pions, kaons and $\eta$ mesons \cite{STAR:2009qzv,PHENIX:2006mhb,STAR:2006xud,PHENIX:2013kod} with the corresponding ALICE $p$Pb measurements \cite{ALICE:2018vhm,ALICE:2016dei,ALICE:2021est} led to a consistent description of the data and to a considerable reduction in the nuclear gluon uncertainty. The fact that consistent global fits down to $p_T > 3 \, {\rm GeV}$  are possible can be also taken as an indication that higher-twist final-state rescattering (cf.\ Sec.\ \ref{sec:2.3}) is indeed a subleading effect. The inclusive hadron data have therefore also been retained in the latest nCTEQ15HQ analysis \cite{Duwentaster:2022kpv}. Alternatively, the RHIC data in DAu collisions have also been interpreted in terms of nuclear-modified FFs \cite{Sassot:2009sh,Zurita:2021kli}, which were used in the DSSZ \cite{deFlorian:2011fp} global analysis of nuclear PDFs and resulted in reduced nuclear effects for the gluon PDF.

The latest LHC measurements of $R_{p{\rm Pb}}$ for high-$p_T$ neutral pion production come from LHCb \cite{LHCb:2022tjh}, complementing the ALICE midrapidity data with forward/backward measurements. While the forward-rapidity data agree with the NLO predictions with nuclear PDFs, there appears to be a slight normalization difference between the predictions and the LHCb data at negative rapidities. The preliminary LHCb data for $\eta$ mesons looks consistent with nuclear PDFs \cite{LHCb:2023iyw}. These data are not yet included in global fits. The LHCb forward pion data also agree well with the LHCb forward charged-hadron ($h^\pm$) data \cite{LHCb:2021vww}. The corresponding backward data are, however, in disagreement with the nuclear PDF predictions, which hints that the baryon production in the lead-going direction cannot be described solely within the factorization. The same issue is visible at midrapidity as well \cite{ATLAS:2022kqu,CMS:2015ved,CMS:2016xef,ALICE:2018vuu} and is more pronounced at RHIC \cite{PHENIX:2003qdw,STAR:2003pjh}. Moreover, it has been noticed that even in simpler $pp$ collisions at LHC energies the collinear factorization around midrapidity appears to be applicable only at $p_{\rm T} \gtrsim 10 \, {\rm GeV}$ for $h^{\pm}$ production  \cite{dEnterria:2013sgr}. As a result, only the production of mesons is considered in global fits of nuclear PDFs.

\subsection{Jets}
\label{sec:5.4}

Jet measurements in $p$Pb collisions probe the intermediate- to large-$x$ regime of nuclear PDFs at large interaction scales ($Q^2 \gtrsim 10^3$ GeV$^2$). A complication in $p$Pb compared to $pp$ collisions is the significantly larger background from the underlying event. Indeed, in Glauber-type models an average $p$Pb collision contains around $7\pm5$ $pN$ interactions \cite{Loizides:2017ack}. To reduce the model dependence in how the multi-parton interactions (MPIs) are dealt with, the jet $p_T$ must therefore be large enough, or the jet cone must be small enough, to reduce the probability of particles from MPIs to occupy the same phase space. However, at too small cone sizes the jet cross sections become unstable due to an incomplete cancellation of infrared divergences. Also the hadronization corrections, which tend to widen the partonic jets, grow. While in $pp$ collisions all non-perturbative corrections are applied to the theoretical predictions, the $p$Pb data have already been subtracted for the backgrounds from MPIs. This works rather well, i.e.\ the obtained ratios $R_{p{\rm Pb}}$ are broadly consistent with the expectations from nuclear PDFs.

\begin{figure}
\includegraphics[width=\linewidth]{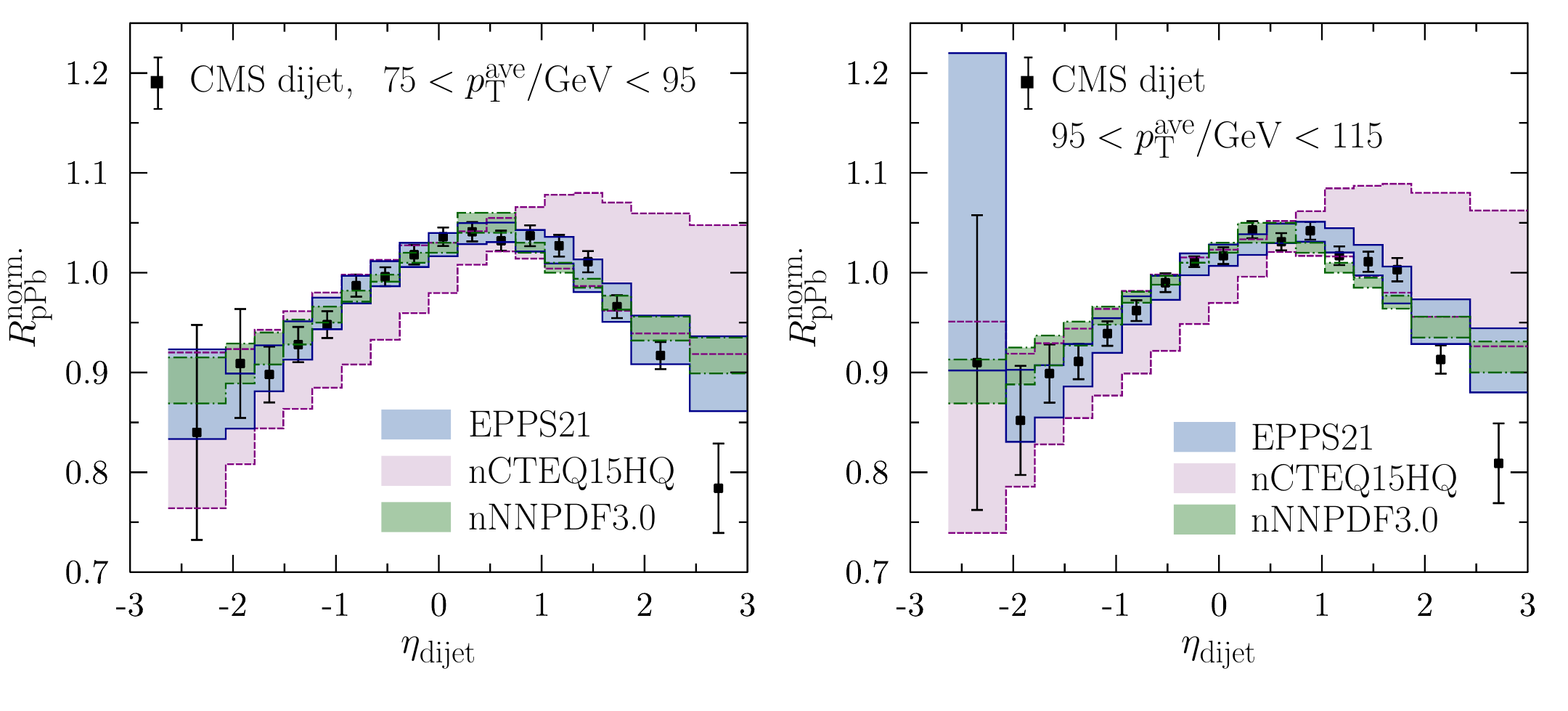} 
\caption{Comparison of the normalized CMS dijet nuclear modification ratios \cite{CMS:2018jpl} with NLO calculations using EPPS21, nCTEQ15HQ and nNNPDF3.0 nuclear PDFs.}
\label{figdijet}
\end{figure}

Currently, the most constraining data are the Run-I CMS dijet data differential in the average $p_T$ ($p^{\rm ave}_{\rm T}$) and rapidity ($\eta_{\rm dijet}$) of the two jets \cite{CMS:2018jpl}. They supersede the earlier dijet data \cite{CMS:2014qvs}, which were included in the EPPS16 analysis. The cross sections are normalized to the rapidity-integrated cross section, so that most of the systematic uncertainties cancel. The resulting spectra in $pp$ are then so precise that they challenge the theoretical description, the NLO perturbative QCD calculations being in tension with the data \cite{Eskola:2019dui}. The impact of the NNLO corrections \cite{Currie:2017eqf} is still unclear. Despite this tension, ratios of normalized cross sections 
between $p$Pb and $pp$ collisions, $R_{\rm pPb}^{\rm norm.}$, are broadly consistent with nuclear PDFs and included in the EPPS21 and nNNPDF3.0 global fits, where they have a large impact on the gluon. However, the most forward data points at the edge of the detector acceptance indicate a suppression, which cannot be fitted. This questions the reliability of extracting nuclear PDFs from these data.

Part of the CMS dijet data are compared in Fig.\ \ref{figdijet} with NLO calculations using EPPS21, nCTEQ15HQ and nNNPDF3.0. The depletion in the backward direction $\eta_{\rm dijet} \ll 0$ indicates a presence of an EMC effect for gluons (though it is partially obscured due to the contribution of valence quarks), whereas the depletion in the forward direction $\eta_{\rm dijet} \gg 0$ is consistent with the expected gluon shadowing. At $\eta_{\rm dijet} \gg 0$, the less shadowed gluons of nCTEQ15HQ seen in Fig.~\ref{fig:nPDFs} lead to the observed higher prediction for $R_{\rm pPb}^{\rm norm.}$. The nCTEQ15HQ error bands are the widest, as these data were not included in the fit. The EPPS21 and nNNPDF3.0 uncertainties are smaller, and near $\eta_{\rm dijet} \sim 1$ there is even a discrepancy among them. Data for single inclusive jets are also available \cite{ATLAS:2014cpa,CMS:2016svx,ALICE:2023ama}, but the uncertainties are clearly larger than in the dijet measurements. 

\subsection{Heavy quarks and quarkonia}
\label{sec:5.5}

\renewcommand{\arraystretch}{1.05}
\begin{table}
\centering
\tabcolsep4.0pt
\caption{Heavy quark production data available from LHC $p$Pb collisions.}
\label{tab:03}
\hspace{-1cm}\begin{tabular}{lcccccccc}
\\
\toprule
{\sc Observable ${\cal O}$}
& $D^0$
& $J/\psi$ 
& $\Upsilon(1S)$
& $\psi(2S)$
& B$^0$, B$^\pm$
& $c$ jet
& $b$ jet
\\
\midrule
\textcolor{blue}{\sc Run-I:}\\
ATLAS & & \cite{ATLAS:2015mpz,ATLAS:2017prf}$^{\rm a}$ & \cite{ATLAS:2017prf}$^{\rm a}$ & \cite{ATLAS:2017prf}$^{\rm a}$ \\
CMS   & & \cite{CMS:2017exb}$^{\rm a}$ & \cite{CMS:2022wfi} & \cite{CMS:2018gbb}$^{\rm a}$ & & \cite{CMS:2016wma} & \cite{CMS:2015gcq} \\
ALICE & \cite{ALICE:2014xjz,ALICE:2016yta,ALICE:2019fhe}$^{\rm a}$ & \cite{ALICE:2013snh,ALICE:2015sru}$^{\rm a}$, \cite{ALICE:2021lmn} & \cite{ALICE:2014ict} & \cite{ALICE:2014cgk}$^{\rm a}$ & & & \cite{ALICE:2021wct} \\
LHCb  & \cite{LHCb:2017yua}$^{\rm a,b,c}$ & \cite{LHCb:2013gmv}$^{\rm a}$ & \cite{LHCb:2014rku} \\
\midrule
\textcolor{blue}{\sc Run-II:}\\
ALICE & & \cite{ALICE:2018mml}$^{\rm a}$, \cite{ALICE:2022zig} & \cite{ALICE:2019qie}$^{\rm a}$ & \cite{ALICE:2020vjy}$^{\rm a}$ \\
LHCb  & \cite{LHCb:2022dmh} & \cite{LHCb:2017ygo}$^{\rm a}$ & \cite{LHCb:2018psc}$^{\rm a}$ & & \cite{LHCb:2019avm} \\
\midrule
\textcolor{blue}{\sc Fixed target:}\\
LHCb & \cite{LHCb:2018jry,LHCb:2022cul} & \cite{LHCb:2018jry,LHCb:2022sxs} & & \cite{LHCb:2022sxs} \\
\bottomrule
\end{tabular}
\begin{tabnote}
$^{\rm a}$ included in nCTEQ15HQ \cite{Duwentaster:2022kpv}; $^{\rm b}$ included in EPPS21 \cite{Eskola:2021nhw}; $^{\rm c}$ included in nNNPDF3.0 \cite{AbdulKhalek:2022fyi}.
\end{tabnote}
\end{table}
\renewcommand{\arraystretch}{1}

The possibilities of constraining the gluon PDF with inclusive heavy-flavor production at the LHC have been actively investigated in $pp$ \cite{Cacciari:2015fta,Gauld:2015yia,Gauld:2016kpd,PROSA:2015yid} as well as in $p$Pb \cite{Duwentaster:2022kpv,Eskola:2019bgf,Kusina:2020dki,Gauld:2015lxa,Kusina:2017gkz} collisions. However, the theoretical approaches vary from one analysis to another, and this is also the case in the fits of nuclear PDFs that include heavy-quark data. The EPPS group uses GM-VFNS calculations \cite{Kniehl:2004fy,Helenius:2018uul,Helenius:2023wkn}, in which heavy quarks are active partons above the mass thresholds, resumming collinear logarithms from the initial- and final-state radiation and thereby matching with the variable-flavor structure of the nuclear PDFs. The nNNPDF group employs an approach which supplements fixed-order calculations with a similar, though less complete resummation of collinear logarithms through parton showers \cite{Nason:2004rx, Frixione:2007vw, Alioli:2010xd}. The nCTEQ15HQ group relies on effective matrix-element fitting \cite{Lansberg:2016deg} introduced in Sec.~\ref{sec:3.1}. It uses the fact that in fixed-order calculations the $gg$ initial state dominates at low $p_T$ and allows fitting both open heavy quark and quarkonium production including the hadronization process, which for quarkonia remains to be fully understood \cite{Andronic:2015wma}. 

The four LHC collaborations have collected a vast data set on $D^0$, $B^0$, $B^\pm$, $J/\psi$, $\Upsilon$ and $\psi'$ mesons (cf.\ Tab.\ \ref{tab:03}), which allow to extend the range in $x_N$ to below $10^{-5}$, i.e.\ more than one (two) order(s) of magnitude lower than LHC electroweak boson (jet) production at scales from $m_c^2$ to $10^3$ GeV$^2$ (cf.\ Fig.\ \ref{fig:xQ2plot}). Including these data even partially, the gluon uncertainties of nCTEQ15HQ, EPPS21, and nNNPDF3.0 have shrunk considerably below $x_N=10^{-2}$ in comparison to their respective predecessors nCTEQ15WZ+SIH \cite{Duwentaster:2021ioo}, EPPS16 \cite{Eskola:2016oht}, and nNNPDF2.0 \cite{AbdulKhalek:2020yuc}. While not included in the current global fits, the CMS collaboration has also measured $c$ \cite{CMS:2016wma} and $b$ jets \cite{CMS:2015gcq}, ALICE $b$ jets \cite{ALICE:2021wct}, LHCb inclusive $B$-meson \cite{LHCb:2019avm} production discussed e.g. in Ref.~\cite{Helenius:2023wkn}, and ALICE heavy-flavor decay electrons \cite{ALICE:2019bfx,ALICE:2023xiu}. First heavy-flavor measurements have also been carried out by LHCb in the fixed-target mode with different nuclei (He, Ar, Ne) \cite{LHCb:2018jry,LHCb:2022cul,LHCb:2022sxs}. This may eventually allow to study the $A$-dependence of nuclear PDFs. 

\begin{figure}
    \centering
    \begin{minipage}{.5\textwidth}
        \centering
        \includegraphics[width=1.0\linewidth]{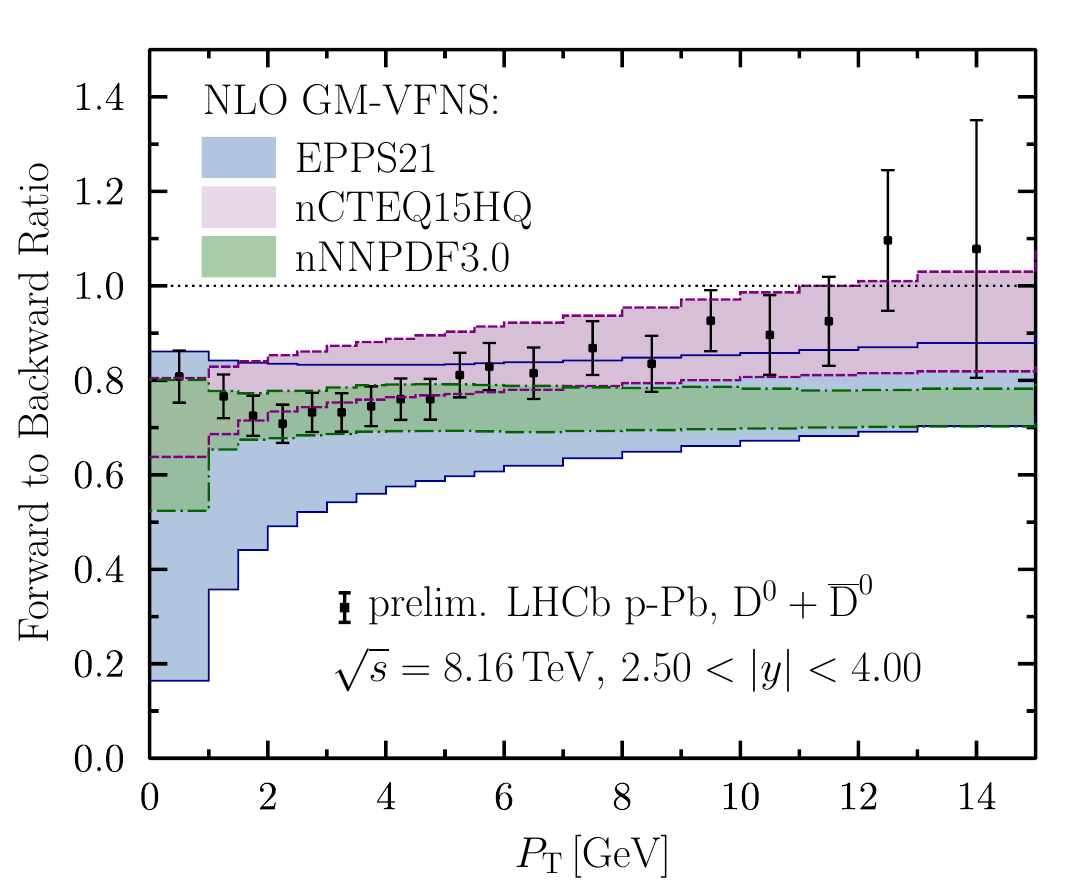}
    \end{minipage}%
    \begin{minipage}{0.5\textwidth}
        \hspace{0cm}\includegraphics[width=1.0\linewidth]{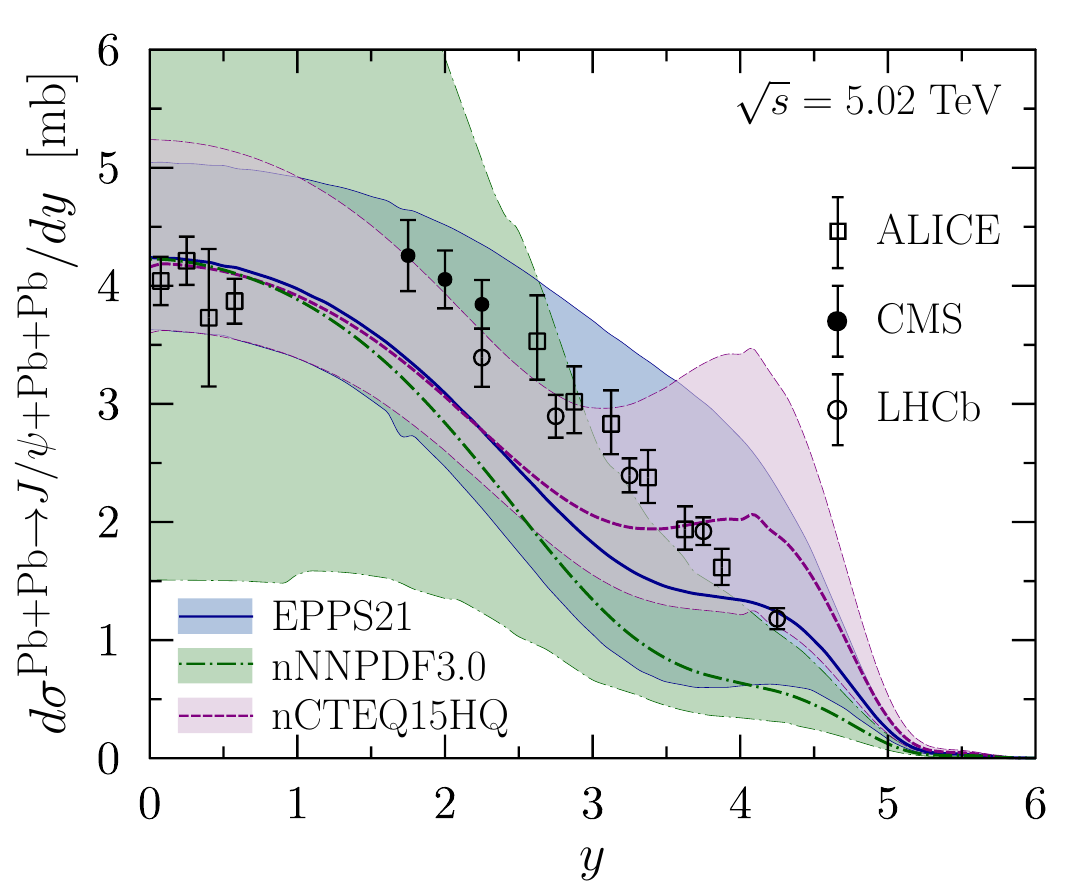}
    \end{minipage}
    \vspace{0.5cm}
        \caption{
        {\bf Left:} The LHCb Run-II $R_{\rm FB}$ \cite{LHCb:2022dmh} compared to NLO GM-VFNS calculations \cite{Helenius:2018uul} using EPPS21, nCTEQ15HQ and nNNPDF3.0 nuclear PDFs. {\bf Right:} The LHC Run-II data \cite{ALICE:2019tqa,ALICE:2021gpt,LHCb:2022ahs,CMS:2023snh} for exclusive $J/\psi$ production in PbPb collisions compared to NLO calculations using EPPS21, nCTEQ15HQ and nNNPDF3.0 nuclear PDFs. The factorization scales $\mu \sim 2.2$ GeV have been chosen to match the ALICE data at $y=0$.}
        \label{figD}
\end{figure}

An important data set in the current global fits is the LHCb Run-I $D^0$ measurement \cite{LHCb:2017yua}, which is included in all three fits. In the forward direction ($y \gg 0$, small $x$), the nuclear modification ratio $R_{p{\rm Pb}}$ shows a clear suppression consistent with shadowing. In the backward direction ($y \ll 0$, larger $x$) at the intersection between shadowing and antishadowing, $R_{p{\rm Pb}}$ is closer to unity. This behavior is consistent with the CMS dijet and $W^\pm$ data. The ALICE $D$-meson data \cite{ALICE:2019fhe} lie at midrapidity in between the LHCb acceptance and have a somewhat different normalization. The recent LHCb Run-II $D^0$ data \cite{LHCb:2022dmh} are consistent with nuclear-PDF predictions in the forward direction (shadowing), but indicate a stronger suppression than expected in the backward direction. Given that these $R_{p{\rm Pb}}$ data use a $pp$ reference interpolated between $5 \, {\rm TeV}$ and $13 \, {\rm TeV}$, $R_{\rm FB}$ could arguably be more accurate. Figure \ref{figD} (left) compares the new LHCb Run-II measurement with the predictions obtained using EPPS21, nCTEQ15HQ and nNNPDF3.0 PDFs in a NLO GM-VFNS calculation \cite{Helenius:2018uul}. Despite the fact that all three use the $5\,{\rm TeV}$ $p$Pb $D^0$ data as an input, there are still significant differences among the predictions. Recently, preliminary LHCb Run-I data on the $R_{p{\rm Pb}}$ of $D^+$ and $D^+_s$ have also appeared \cite{LHCb:2023kqs}. They are consistent with the $D^0$ results at $y \gg 0$, but the $D^+$ data deviate from the $D^0$ results at $y \ll 0$.

The prospects of using top quark production in $p$Pb and PbPb collisions to understand nuclear PDFs were first quantitatively discussed in Ref.~\cite{dEnterria:2015mgr}. While the large mass of the top quark renders the production cross sections small in comparison to charm or beauty production, the process was predicted to be visible at the LHC. Total top quark cross sections have thereafter been measured by CMS \cite{CMS:2017hnw} and ATLAS \cite{2702726} and also in PbPb collisions by CMS \cite{CMS:2020aem}. The ATLAS measurement in $p$Pb is consistent with the nCTEQ15HQ, EPPS21 and TUJU21, but not with the nNNPDF3.0 nuclear PDFs. 

\subsection{Exclusive and inclusive observables in ultraperipheral collisions}
\label{sec:5.6}

Ultraperipheral collisions (UPCs) of ions are interactions in which the approaching nuclei do not touch. Instead, they interact at a distance due to their strong electromagnetic fields \cite{Bertulani:2005ru,Baltz:2007kq}. In comparison to typical minimum-bias $p$Pb (let alone PbPb) collisions, much fewer background processes take place, and the signal processes are thus easier to isolate.

The exclusive production of $J/\psi$ mesons in UPCs has triggered particular interest. The process is dominated by the exchange of an almost real photon. In photon-nucleus collisions, the PDFs appear already at the level of the matrix element
\begin{equation}
\mathcal{M}(\gamma + A \rightarrow J/\psi + A) \sim T_g \otimes f_g^{A} + \sum_q T_{q} \otimes f_q^{A} \,.
\end{equation}
When squared to obtain a cross section, the latter becomes extremely sensitive to PDFs. Several LO studies have been performed in the past \cite{Adeluyi:2011rt,Adeluyi:2012ph,Guzey:2013qza,Guzey:2013xba}, but the first NLO calculations for PbPb collisions have appeared only very recently \cite{Eskola:2022vpi,Eskola:2022vaf} despite the fact that the NLO coefficient functions $T_{g,q}$ have been known for some time \cite{Ivanov:2004vd}. Figure \ref{figD} (right) compares NLO calculations with several recent nuclear PDFs with the combined experimental data from the LHC \cite{ALICE:2019tqa,ALICE:2021gpt,LHCb:2022ahs,CMS:2023snh}. The factorization scales have been chosen to match the ALICE data at mid-rapidity. While the central theory values do not reproduce the behavior of the data particularly well, the nuclear PDF error bands are much wider than the data uncertainties. This indicates that these data should further constrain the nuclear PDFs. Unfortunately, the process is perturbatively unstable: at LO only gluons contribute, while at NLO there is also a contribution from the quark singlet, which can even dominate at NLO. The reason is that the LO and NLO gluon contributions enter $T_g$ with opposite signs and there is a significant cancellation between the two. It has been argued that the theoretical uncertainties could be brought under better control by summing logarithmically enhanced contributions at small $x$ \cite{Jones:2015nna} and by considering power corrections in the coefficient functions \cite{Jones:2016ldq}. There are also other theoretical uncertainties associated e.g.\ with modeling of the photon flux and how nuclear generalized PDFs (GPDs) and collinear PDFs are related \cite{Dutrieux:2023qnz}. 

In addition to exclusive observables, also inclusive processes such as dijet photoproduction are sensitive to nuclear PDFs. Here a photon emitted from one nucleus breaks up the other to produce a hadronic final state that contains two hard jets. This process has been calculated in LO with parton showers  \cite{Helenius:2018mhx} and in NLO \cite{Guzey:2018dlm}. Preliminary measurements by ATLAS also exist \cite{ATLAS:2017kwa,ATLAS:2022cbd}, and the 2017 measurement qualitatively agrees with the NLO calculation. However, the ATLAS data make use of forward neutrons and rapidity gaps to resolve the photon-going direction on an event-by-event basis, and this excludes the diffractive component of UPCs, which is included in NLO calculations with standard PDFs \cite{Guzey:2020ehb}. The imposed experimental conditions also require further modeling associated with preventing a Coulomb break-up of the photon-emitting nucleus and with finite-size effects.

\section{Other developments}
\label{sec:06}

\subsection{Electron Ion Collider}
\label{sec:6.1}

The Electron Ion Collider (EIC) currently under construction at BNL will extend the kinematic region of lepton-nucleus scattering compared to fixed target experiments by one order of magnitude in $x$ (to a few times 10$^{-4}$) and $Q^2$ (to $10^3$ GeV$^2$). This region is of course still considerably smaller than the one accessible at the LHC (cf.\ Fig.\ \ref{fig:xQ2plot}), but the environment will be much cleaner than in $pA$ collisions, making it easier to disentangle cold nuclear matter (i.e.\ leading twist, factorizable) effects from other possible (higher twist, non-factorizable) contributions. In addition, the EIC will allow for measurements on a large variety of nuclei in NC and CC DIS as well as in photoproduction and with inclusive, semi-inclusive (e.g.\ identified hadron) and exclusive (e.g.\ diffractive) final states \cite{Accardi:2012qut,AbdulKhalek:2021gbh}.

While $F_2^A$ is sensitive to the momentum distributions of gluons mainly through scaling violations (cf.\ Eq.\ \ref{eq:35}), the EIC will also allow for measurements of the longitudinal structure function \cite{Altarelli:1978tq,Bandeira:2022dvu}
\beq
 F_L^A(x,Q^2)=\frac{\alpha_s(Q^2)}{2\pi}x^2\int_x^1 \frac{dz}{z^3}\le \frac{8}{3} F_2(x,Q^2)+4\sum_q e_q^2\lr1-\frac{x}{z}\rr z f_g^A(x,Q^2)\re,
\eeq
which has a direct contribution from gluons. Measuring $F_L^A$ will, however, require operation at different center-of-mass energies. The impact of inclusive NC DIS measurements at the EIC on global fits of nuclear PDFs has been investigated in Refs.~\cite{AbdulKhalek:2021gbh,Aschenauer:2017oxs,Khalek:2021ulf} in the frameworks of nCTEQ15WZ, EPPS16 and nNNPDF2.0 with several nuclei and beam-energy configurations (see also Ref.\ \cite{AbdulKhalek:2019mzd}). At low $Q^2$, the predicted impact is significant for all partonic flavors other than the strange quark. At higher $Q^2$, the better constrained gluon also leads to a better constrained strange quark PDF. Charm tagging allows to access the reduced charm cross section
\beq
 \sigma^{c\bar{c}}_{\rm red} = \frac{d^2\sigma^{c\bar{c}}}{dxdQ^2}\frac{xQ^4}{2\pi\alpha^2[1+(1-y)^2]}=F_2^{c\bar{c}}-\frac{y^2}{1+(1-y)^2}F_L^{c\bar{c}}
\eeq
related to the charm structure functions $F_{2,L}^{c\bar{c}}$, sensitive to the gluon and a potential intrinsic charm content in the nucleon, and thus to further reduce the uncertainties \cite{Aschenauer:2017oxs,Kelsey:2021gpk}. Dijets in DIS \cite{Klasen:2017kwb,Arratia:2019vju} and photoproduction \cite{Klasen:2018gtb,Aschenauer:2019uex,Guzey:2020zza} as well as charm jets \cite{Arratia:2020azl} provide further information, the latter in particular on the strange quark in CC DIS. Similarly to the LHC, exclusive vector meson production is highly sensitive to the gluon \cite{Chen:2018vdw,Goncalves:2020ywm}. Diffractive final states with a large rapidity gap or identified hadrons in the forward direction will allow to access the completely unknown territory of diffractive nuclear PDFs \cite{Guzey:2020gkk,Deak:2020mlz}.

Taken together, the EIC measurements will allow for a greater parametric flexibility in the $x$-dependence at the starting scale $Q_0^2$ similar to proton PDFs, lead to a more reliable $A$-dependence, make parameterizations possible not just in $A$, but also in $Z$, and allow to move away from the nuclear stability line and to study mirror nuclei. The EIC should also help to answer the question of potential different nuclear effects in CC and NC DIS, shed light on shadowing, gluon saturation \cite{Marquet:2017bga,Mantysaari:2019hkq,Tong:2022zwp}, transverse momentum distributions (TMDs) \cite{Dumitru:2018kuw,Caucal:2023nci}, the transition to the color glass condensate (CGC) \cite{Caucal:2021ent,Liu:2023aqb} as well as the EMC effect across a wide range of $A$ and energy scales. With polarized beams of light nuclei ($^2$H, $^3$H, $^3$He), even the polarized EMC effect could be investigated \cite{Accardi:2012qut,AbdulKhalek:2021gbh}.

\subsection{Lattice QCD}
\label{sec:6.2}

In the non-perturbative lattice QCD approach, the four-dimensional space-time is discretized and QCD regularized on a finite Euclidean lattice. Correlation functions are then computed numerically in the path integral formalism using methods adapted from statistical mechanics, and the results are extrapolated to the continuum and infinite volume limits. To make contact with experimental data, lattice QCD calculations must demonstrate control over all sources of systematic uncertainty including discretization effects, extrapolation from unphysical pion masses, finite-volume effects, and renormalization of composite operators.

Light-cone quantities like PDFs cannot be calculated directly on a Euclidean lattice. Instead, the traditional approach has been to determine the matrix elements of local twist-two operators that can be related to the Mellin moments of PDFs. In principle, given a sufficient number of Mellin moments, PDFs can be reconstructed from the inverse Mellin transform. In practice, however, the calculation is limited to the lowest three moments, since power-divergent mixing occurs between twist-two operators. Three moments are insufficient to fully reconstruct the momentum dependence of the PDFs without significant model dependence. The lowest three moments do provide, however, useful information both as benchmarks of lattice calculations and as constraints in global extractions of PDFs \cite{Lin:2017snn}.

Direct extractions of the $x$ dependence of PDFs have been attempted based on quasi-PDFs in large-momentum effective theory \cite{Ji:2020ect}, pseudo-PDFs \cite{Radyushkin:2016hsy} and other methods. Quasi-PDFs are defined as Fourier transforms of the matrix elements, whereas pseudo-PDFs are transforms in Ioffe time. For positive ($u-d$) and negative ($\bar u-\bar d$) isovector quark combinations, the quasi-PDFs at the physical pion mass were found to agree with global fits at large $x>0.1$ and $x>0.4$, respectively \cite{Lin:2017ani}. Gluon quantities are much noisier than quark disconnected loops and require calculations with very high statistics. Up to perturbative matching and power corrections, the Fourier transform of the gluon quasi-PDF was found to be compatible with the one of global fits within the statistical uncertainty \cite{Fan:2018dxu}. Calculating the small-$x$ behavior requires larger boost momenta, as this results in a faster decay of the matrix elements, so that truncations in the Fourier transform matter less \cite{Constantinou:2020hdm}. Fourier transforms of strange and charm quark PDFs have also been obtained, the former being about five times larger and both being smaller than those in global fits, possibly due to missing contributions from other flavor distributions. A full analysis of lattice QCD systematics must still be performed. Nevertheless, the strangeness asymmetry ($s-\bar s$) in the region of $0.3 < x < 0.8$ was found to be very small with high precision compared to the uncertainty in global fits \cite{Zhang:2020dkn}. Including these lattice data in a global fit therefore greatly reduces the size of the $s-\bar s$ error band in the large-$x$ region \cite{Hou:2022onq}.

Lattice QCD studies of nuclear structure are currently restricted to low $A$ and unphysical pion masses. In particular, the ratio of the longitudinal momentum fraction carried by the positive isovector quark combination in $^3$He to the one in the free nucleon was found to be consistent with unity at the few-percent level. This is in agreement with, but more precise than, current determinations from global fits. Including this lattice result in the nNNPDF2.0 global fit framework reduces the uncertainty on the isovector momentum fraction ratio by a factor of 2.5 and leads to a more precise extraction of the $u$ and $d$ quark distributions in $^3$He \cite{Detmold:2020snb}. Previously, the nuclear modification of the gluon momentum fraction for $A\leq3$ was found to be less than $\sim$10\%. This is consistent with expectations from phenomenological quark distributions and the momentum sum rule \cite{Winter:2017bfs}.

\subsection{Relations to other phenomena}
\label{sec:6.3}

The global analysis of nuclear PDFs outlined in this review indicates that, at sufficiently high interaction scales, collinear factorization is a consistent way to describe 
lepton-nucleus and proton-nucleus collisions. However, towards low interaction scales there are theoretical reasons to believe that the role of higher-twist effects, suppressed by inverse powers of the interaction scale, become increasingly important in comparison to simpler lepton-proton and proton-proton collisions. In the parton model, these effects arise from processes where two initial-state partons recombine \cite{Gribov:1983ivg,Mueller:1985wy,Zhu:1998hg,Zhu:1999ht}. In general, these processes modify the linear DGLAP evolution by slowing down the evolution at small $x$ and increasing it at intermediate $x$. Since the spatial density of partons is higher in large nuclei, the effect should be more pronounced there and lead to dynamically generated shadowing and antishadowing \cite{Rausch:2022nkv}. At very low $Q^2$, the recombination of an even larger number of partons becomes eventually important and, when resummed, can be interpreted as saturation \cite{Gelis:2010nm,Morreale:2021pnn}. Finding a conclusive signature of saturation has proven rather difficult, calculations based on nuclear PDFs and the saturation picture giving very similar results. Even forward $D$-meson production in $p$Pb collisions, which probes the nucleus down to $x \sim 10^{-5}$ at nearly non-perturbative interaction scales, has not revealed clear deviations from the linear DGLAP dynamics. In addition, the resummation of small-$x$ BFKL logarithms in the language of PDFs also have a tendency to slow down the linear DGLAP evolution \cite{Bonvini:2016wki,Ball:2017otu}, which further complicates the search for true saturation effects.

There is a heated discussion revolving around the question whether in very central, high-multiplicity $p$Pb collisions -- and even in smaller systems like $pp$ or $\gamma A$ collisions -- one creates a state of matter that has collective liquid-type properties -- a droplet of Quark-Gluon plasma (QGP) \cite{Pasechnik:2016wkt}. Some characteristic features have been experimentally observed \cite{Nagle:2018nvi}, which typically consist of correlations between particles. However, the measurements can also be interpreted in terms of the initial-state geometry \cite{Schlichting:2016sqo}. Furthermore, the way in which final-state QCD particles hadronize into a color-neutral state has been shown to display features that could accidentally be attributed to a liquid-like behavior \cite{OrtizVelasquez:2013ofg}. It should also be kept in mind that jet quenching has not been observed in $p$Pb collisions, which challenges the QGP picture of small systems and is a concrete difference between observations in $p$Pb and PbPb collisions. Multiplicity-integrated $p$Pb cross sections appear to be consistent with collinear factorization and process-independent nuclear PDFs. The relations between the observations discussed above remain open questions at this moment. 

In the case of heavy-ion collisions, the formation of a QGP is nowadays a generally accepted phenomenon. Nevertheless, even in heavy-ion collisions the LHC data for electroweak boson \cite{ALICE:2020jff,ALICE:2022cxs,ATLAS:2019ibd,ATLAS:2019maq} and high-$p_T$  direct photon production \cite{ATLAS:2015rlt,CMS:2012oiv,PHENIX:2012jbv} are consistent with collinear factorization and process-independent nuclear PDFs \cite{Eskola:2020lee}. Thus, there is no reason to believe that the initial state of heavy-ion collisions would not be dictated by nuclear PDFs. This idea has been pursued in the Eskola-Kolhinen-Ruuskanen-Tuominen model of heavy-ion collisions, whose most recent versions \cite{Niemi:2015qia} apply NLO perturbative QCD calculations and impact-parameter dependent nuclear PDFs \cite{Helenius:2012wd} to compute the inital conditions for the subsequent fluid-dynamical evolution of the system.

The nuclear PDFs also find use in the field of neutrino astronomy and cosmic-ray physics \cite{Bhattacharya:2016jce,Bertone:2018dse,Reno:2023sdm}. Interactions of neutrinos coming from outer space can be measured in large neutrino telescopes such as IceCube, KM3NeT and Baikal, where the neutrinos interact with water or ice. Precise theoretical calculations of the cross sections require nuclear PDFs as an input. In addition, protons from astrophysical sources can collide with the air molecules in the atmosphere, which can also produce neutrinos. Precise calculations of the cross sections for these secondary neutrinos require nuclear PDFs as well. 

\section{Conclusion}
\label{sec:07}

During the last 25 years of research in nuclear PDFs, the field has undergone an enormous development. Methodologically, simple fits, performed by eye at LO, have matured into rigorous statistical analyses, including machine-learning techniques, at NLO and NNLO with full error estimates. Nonetheless, global nuclear-PDF analyses are still driven by experimental measurements, and in this respect the $p$Pb collisions carried out during the past decade at the LHC have opened up a wide, previously unexplored regime both in terms of kinematics and processes. The following items summarize our take-home messages:

\begin{summary}[SUMMARY POINTS]
\begin{enumerate}
\item 
Despite the theoretical and experimental advances, there are still significant differences among the independent global analyses of nuclear PDFs, both in terms of the extracted nuclear modifications of PDFs as well as the absolute nuclear PDFs. In several places the central values of a given analysis can be outside the error bands of the others, and in some cases even the error bands of two given analyses do not overlap. The widths of the error bands can be also very different. The most significant factors behind the observed differences can be attributed to (i) the assumed form of the non-perturbative parameterization of nuclear PDFs at low $Q^2$, (ii) the data selection, (iii) fitting absolute cross sections vs. ratios of cross sections, (iv) the theoretical treatment of heavy flavors, and (v) different baseline free-proton PDFs. As highlighted in Figs.~\ref{fig:nPDFs}--\ref{figD}, these differences also lead to visible effects in observables. This underscores the need to carry out the global analysis independently in several groups. To faithfully chart the theoretical uncertainties in quantities that depend on nuclear PDFs, it is thus recommended to use more than one set of nuclear PDFs. 

\item 
At the moment, NLO accuracy is the standard in the field of nuclear PDFs, the full NNLO accuracy being limited by the existence or public availability of cross-section codes. Most $pA$ data can be well described within the NLO calculations, but in the case of some observables there is a confirmed (DY below the $Z$ peak) or conjectured (prompt photons, jets) need for NNLO QCD. The NNLO accuracy also reduces the theoretical uncertainties in observables sensitive to effects of partonic saturation, offers a standard candle to Glauber modeling of heavy-ion collisions through precise predictions for electroweak observables, and should lead to more precise predictions for astrophysical applications.

\item
In the long run, the global analysis of nuclear PDFs should be extended to include the proton and the deuteron. As of today, most of the free-proton fits still utilize heavy-target -- particulary neutrino DIS -- data to constrain the full flavor decomposition. These same data can be (and are), however, also taken as constraints on the nuclear modifications of PDFs. To fully chart the interplay between the two requires a simultaneous extraction of the free-proton and nuclear PDFs.

\end{enumerate}
\end{summary}

\section*{DISCLOSURE STATEMENT}

The authors are not aware of any affiliations, memberships, funding, or financial holdings that might be perceived as affecting the objectivity of this review. 

\section*{ACKNOWLEDGMENTS}

The authors thank their nCTEQ and EPPS colleagues for their collaboration and useful discussions, E.~Nocera for providing the nNNPDF3.0 values for Fig.~7, and V.~Guzey for providing the nCTEQ15HQ values for the right-hand panel of Fig.~8. MK thanks his ALICE colleagues for their collaboration and acknowledges funding by the BMBF through project 05P21PMCAA and DFG through GRK 2149 and SFB 1225 “Isoquant,” project-id 273811115. HP acknowledges funding by the Academy of Finland through the Center of Excellence in Quark Matter, project 346326. The results shown in Fig.~6 and in the left-hand panel of Fig.~8 have been computed for this review using computing resources of the Finnish IT Center for Science (CSC), project jyy2580.


\clearpage
\section*{Supplemental Material}
\label{sect:Supplementarymaterial}

\begin{figure}[h!]
\includegraphics[width=1.0\textwidth]{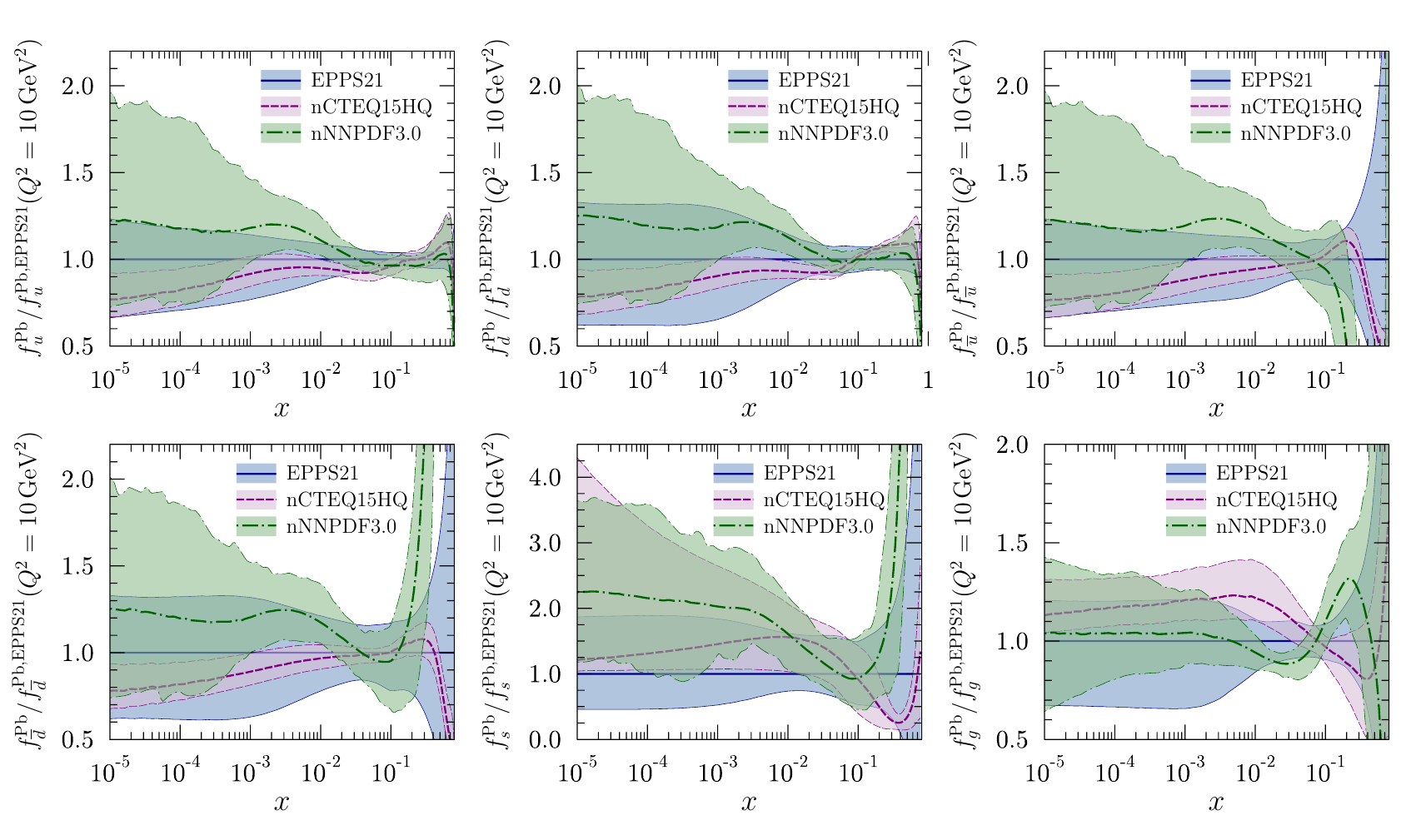}
\caption{Comparison of the $^{208}$Pb nuclear PDFs resulting from the EPPS21 (full, blue) \cite{Eskola:2021nhw}, nCTEQ15HQ (dashed, red) \cite{Duwentaster:2022kpv} and nNNPDF3.0 (dot-dashed, green) \cite{AbdulKhalek:2022fyi} global analyses at $Q^2 = 10\,{\rm GeV}^2$, normalized to the central values of EPPS21. The uncertainty bands correspond to 90\% CL.}
\label{fig:absnPDFs}
\end{figure}

\end{document}